\documentclass[%
superscriptaddress,
5groupedaddress,
preprint,
aps,
pra,
superscriptaddress
]{revtex4-2}
\usepackage{amsthm, amssymb, amsmath}
\usepackage[english]{babel}
\usepackage{natbib} 
\usepackage{graphicx}
\usepackage{dcolumn}
\usepackage{bm}
\usepackage{braket}
\usepackage{tikz}
\usepackage{hyperref}
\usepackage[caption=false]{subfig}
\usepackage[version=4]{mhchem}
\usetikzlibrary{quantikz2}
\usepackage{multirow}
\usepackage{makecell}
\usepackage{soul}
\usepackage{float}
\usepackage{appendix}
\usepackage[table,xcdraw]{xcolor}
\usepackage{cleveref}
\usepackage{booktabs}

\usepackage{algorithm}
\usepackage{algpseudocode}
\usepackage{float} 

\usepackage{xr}
\usepackage{acro}

\DeclareAcronym{vqe}{
  short = VQE,
  long  = Variational Quantum Eigensolver,
}

\DeclareAcronym{adaptVQE}{
  short = ADAPT-VQE,
  long  = Adaptive Derivative-Assembled Pseudo-Trotter Variational Quantum Eigensolver,
}

\DeclareAcronym{ggaVQE}{
  short = GGA-VQE,
  long  = Greedy Gradient-Free Adaptive VQE,
}

\DeclareAcronym{bch}{
  short = BCH,
  long  = Baker-Campbell-Hausdorff,
}

\DeclareAcronym{ucc}{
  short = UCC,
  long  = Unitary Coupled Cluster,
}

\DeclareAcronym{ups}{
  short = UPS,
  long  = Unitary Product State,
}

\DeclareAcronym{fci}{
  short = FCI,
  long  = Full Configuration Interaction,
}

\DeclareAcronym{1-cf}{
  short = 1-CF,
  long  = one-parameter cost function,
}

\DeclareAcronym{qeb}{
  short = QEB,
  long  = Qubit Excitation Based,
}

\DeclareAcronym{ceo}{
  short = CEO,
  long  = Coupled Exchange Operator,
}

\renewcommand{\selectlanguage}[1]{}
\begin{document}

\newcommand{\bea}{\begin{eqnarray}}
\newcommand{\eea}{\end{eqnarray}}
\newcommand{\EQ}[1]{Equation~(\ref{#1})} %
\newcommand{\eq}[1]{Eq.~(\ref{#1})} %
\newcommand{\eqs}[1]{Eqs.~(\ref{#1})} %
\newcommand{\fig}[1]{Fig.~\ref{#1}} %
\newcommand{\figs}[1]{Figs.~\ref{#1}} %
\newcommand{\CR}[1]{\hat a^{\dagger}_{#1}}
\newcommand{\AN}[1]{\hat a_{#1}}

\newcommand{\LA}[1]{\mathfrak{#1}}
\newcommand{\HG}{\hat G}

\newcommand{\BC}[1]{\hat \beta^{\dagger}_{#1}}
\newcommand{\BA}[1]{\hat \beta_{#1}}
\newcommand{\EU}[1]{\hat E^{#1}}
\newcommand{\ED}[1]{\hat E_{#1}}
\newcommand{\EE}[2]{\hat E_{#2}^{#1}}
\newcommand{\EX}[2]{\mathcal{E}_{#2}^{#1}}
\newcommand{\GP}[1]{\hat \gamma_{#1}}
\newcommand{\kh}{\hat \kappa}
\newcommand{\OO}{\hat O}
\newcommand{\HC}{\hat C}
\newcommand{\HA}[1]{\hat A_{#1}}


\newcommand{\bfr}{\mathbf{r}} %
\newcommand{\E}{\textrm{e}} %
\newcommand{\I}{\mathrm{i}\mkern1mu} %
\newcommand{\RR}{\mathbf{R}} %
\newcommand{\HP}{\hat P} %
\newcommand{\rr}{\mathbf{r}} %
\newenvironment{NB}{\color{red}NB: }{\ignorespacesafterend}  %
\newcommand{\HH}{\hat H} %
\newcommand{\HU}{\hat U} %
\newcommand{\hz}{\hat z} %

\newcommand{\AFI}[1]{\textcolor{blue}{[AFI: #1]}}
\newenvironment{KMZ}{\par\begingroup\color{orange}}{\par\endgroup}

\title{Resource-efficient energy-based operator selection in fermionic ADAPT-VQE via exact Hamiltonian transformation}

\author{Emanuele Rossi}
\email{emaro@kemi.dtu.dk}
\affiliation{DTU Chemistry, Technical University of Denmark, Kemitorvet 207, 2800 Kongens Lyngby, Denmark}
\author{Erik Rosendahl Kjellgren}
\affiliation{Department of Physics, Chemistry and Pharmacy, University of Southern Denmark, Campusvej~55, DK--5230 Odense M, Denmark}
\author{Artur F. Izmaylov}
\affiliation{Department of Physical and Environmental Sciences,
University of Toronto Scarborough, Toronto, Ontario M1C 1A4, Canada}
\affiliation{Chemical Physics Theory Group, Department of Chemistry, University of Toronto, Toronto, Ontario M5S 3H6, Canada}
\author{Stephan P.A. Sauer}
\affiliation{Department of Chemistry, University of Copenhagen, Universitetsparken 5, 2100 Copenhagen,
Denmark}
\author{Karl Michael Ziems}
\email{K.M.Ziems@soton.ac.uk}
\affiliation{School of Chemistry, University of Southampton, Highfield, Southampton SO17 1BJ, United Kingdom}
\affiliation{DTU Chemistry, Technical University of Denmark, Kemitorvet 207, 2800 Kongens Lyngby, Denmark}
\author{Sonia Coriani}
\affiliation{DTU Chemistry, Technical University of Denmark, Kemitorvet 207, 2800 Kongens Lyngby, Denmark}

\date{\today}

\begin{abstract}
Energy-based variants of ADAPT-VQE select operators by minimizing a one-parameter cost function for each candidate generator. In fermionic implementations, the cost of reconstructing this function often becomes a bottleneck. We reduce this cost using exact fermionic Hamiltonian transformations to obtain a generator-dependent partition of the Hamiltonian that reconstruct the same one-parameter energy function from fragment measurements. This method reproduces the fermionic Rotoselect scoring function exactly with approximately half the Pauli-string measurements. We use this formulation to benchmark the gradient-based and energy-based selection approaches in combination with two ansatz-optimization strategies---\textit{last}, where only the appended operator is optimized, and \textit{full}, where the full ansatz is re-optimized---and with both fixed-orbital and orbital-optimized formulations. The benchmark comprises LiH, \ce{BeH2}, and \ce{H2O} at both equilibrium and stretched geometries. In the most weakly correlated system, pairing energy-based selection with \textit{last} optimization enables the resource-efficient construction of an accurate ansatz, which avoids any VQE optimization. As correlation increases, full ansatz re-optimization and orbital optimization become the main factors governing convergence and overall resource cost. This study shows how exact Hamiltonian transformations provide an effective route to reducing the measurement overhead of fermionic energy-based ADAPT-VQE. Moreover, the benchmark clarifies the relative role of operator selection approach, parameter reoptimization strategy, and orbital optimization in the performance of ADAPT-VQE.
\end{abstract}

\maketitle

\section{Introduction}
Adaptive variants of the \ac{vqe} are among the most promising approaches for quantum chemistry on near-term quantum hardware \cite{Peruzzo_2014,Cao_2019,ADAPT_2019}. They are based on an iterative approach to ansatz construction, which alternates a sequence of two steps: \textit{selection}, wherein an anti-Hermitian generator is selected from a pool and its exponential is appended to the ansatz, and \textit{optimization}, wherein the extended ansatz is optimized according to the \ac{vqe} procedure. The adaptive protocol produces a compact, system-specific ansatz and is reported to mitigate some issues associated with ansatz optimization connected to rough variational landscapes and barren plateaus \cite{Grimsley_2023}.

Recent developments in adaptive \ac{vqe} have addressed several complementary challenges: operator-pool design and ansatz parametrization~\cite{ADAPT_2019,qubit_ADAPT_2021,Yordanov_2021,Ramoa_2025}; measurement overhead in generator selection~\cite{Anastasiou_2024,Long_Dalton_2024,Nykanen_2025,Anastasiou_Mayhall_2023,Shkolnikov_2023}; global optimization and operator reordering strategies \cite{Burton_DISCO_2023}; reduction of the optimization cost \cite{Ramôa_Santos_Economou_2025,Tang_Economou_2024}; application of optimization and ansatz pruning and re-ordering strategies to avoid local minima and gradient troughs~\cite{Stadelmann_2025,Lan_Liang_2022,Vaquero-Sabater_2025}. 

Within this broader context, the choice of how candidate generators are scored remains especially important because it affects both the measurement cost of each ADAPT iteration and the quality of the ansatz. The original \acs{adaptVQE} algorithm uses a gradient-based strategy \cite{ADAPT_2019}, while later variants---based on the Rotoselect algorithm~\cite{Ostaszewski_2021}---adopt an energy-based strategy. In fermionic \ac{adaptVQE}, the repeated scoring of a large operator pool can dominate the quantum cost of the algorithm, making the operator-selection step a central bottleneck \cite{ADAPT_2019,Ramoa_2025}. This work focuses on reducing the measurement overhead associated with energy-based selection strategy.

The Rotoselect algorithm is based on the well-known analytic trigonometric structure of the \ac{1-cf} corresponding to each operator~\cite{Vidal_2018,Parrish_2019,Nakanishi_2020,Ostaszewski_2021,Wierichs_2022}. Finding the minimum of the \ac{1-cf} yields both the corresponding parameter and the operator's energy score~\cite{Ostaszewski_2021}. In an adaptive setting, the parameter value can be used to provide an informed initialization of the appended operator for the subsequent ansatz optimization \cite{ExcitationSolve_2025,GGA_ADAPT_2025}.

The reconstruction of the \ac{1-cf} is based on the parameter-shift rule \cite{Wierichs_2022}. This requires multiple energy evaluations at different parameter values, the number of which depends on the number of eigenvalues of the generator. The cost of these evaluations can quickly exceed those related to the energy gradient, aggravating the measurement overhead associated with the adaptive algorithms. 

In this work, we reduce the measurement overhead of energy-based selection by introducing a resource-efficient fermionic implementation of the existing Rotoselect criterion. Our construction exploits the closed-form theory of fermionic unitary transformations developed by \citeauthor{Evangelista_Magoulas_2025}~\cite{Evangelista_Magoulas_2025} to rewrite the \ac{1-cf} in a generator-dependent fragmented form. The proposed reformulation leaves the asymptotic scaling unchanged but reduces the cost of reconstructing the same Rotoselect \ac{1-cf} by approximately a factor of two. Within the scope of our Pauli-string cost metric, our resource-efficient Rotoselect method reduces the cost of energy-based selection to the level of gradient-based selection. This does not imply cost equivalence of the two methods after cost-reduction measures, such as grouping of commuting Pauli strings, are applied.

Having reduced the measurement cost overhead of energy-based selection, we then investigate the importance of the selection rule relative to other algorithmic choices. For this purpose, we benchmark gradient-based and energy-based selection in combination with two parameter-update strategies---optimization 
of the last appended operator only,
(\textit{last}) 
or full re-optimization of the ansatz (\textit{full})---and with both fixed-orbital and orbital-optimized formulations~\cite{Sokolov_2020,Mizukami_2020,OO_ADAPT_VQE_2024}. 
We carry out the benchmark considering LiH, \ce{BeH2}, and \ce{H2O} in both their equilibrium and stretched geometries, which allows us to contrast weakly and more strongly correlated regimes.

The paper is organized as follows: in Section \ref{sec:adaptive_algs}, we introduce the theory behind adaptive algorithms, briefly reviewing the gradient-based and energy-based selection strategies; in Section \ref{sec:RSe_alg}, we present our resource-efficient Rotoselect method and the details of its implementation; in Section \ref{sec:results}, we start by comparing the selection-cost of gradient-based ADAPT, standard fermionic Rotoselect, and our resource-efficient Rotoselect implementation. We then present the computational benchmark, where we analyze how selection, re-optimization strategy, and orbital-optimization affect overall performance. Section \ref{sec:conclusions} summarizes our conclusions.

\section{Theory}
\subsection{Adaptive algorithms}\label{sec:adaptive_algs}
The energy of the ground state wave function, $\ket{\Psi(\bm{\kappa},\bm{\theta})}$, is given by
\bea\label{eqn:oo_energy_cost_function}
    E(\bm{\theta},\bm{\kappa}) = \bra{\Psi_{\text{ref}}}\hat{U}^{\dagger}(\bm{\theta})\hat{U}^{\dagger}(\bm{\kappa})\hat{H}\hat{U}(\bm{\kappa})\hat{U}(\bm{\theta})\ket{\Psi_{\text{ref}}}~,
\eea
where $\ket{\Psi_{\text{ref}}}$ represents the reference wave function (which we choose to be the Hartree-Fock Slater determinant), $\bm{\theta}$ is the vector of real-valued ansatz parameters, and $\bm{\kappa}$ is the vector of orbital rotation coefficients. The electronic Hamiltonian $\hat{H}$ can be written in second quantization as
\bea\label{eqn:electronic_H}
    \hat{H} = \sum_{pq}h_{pq}\hat{E}_{pq} + \frac{1}{2}\sum_{pqrs}g_{pqrs}\hat{e}_{pqrs}~,
\eea
where $\hat{E}_{pq} = \hat{a}^\dagger_{p\alpha}\hat{a}_{q\alpha}+
\hat{a}^\dagger_{p\beta}\hat{a}_{q\beta}$ is the singlet one-electron excitation operator (with $\alpha$ and $\beta$ indicating the spin orientation corresponding to $m_s=1/2$ and $m_s=-1/2$, respectively), and
\bea\label{eqn:2e_exc_operator}
    \hat{e}_{pqrs} = \hat{E}_{pq}\hat{E}_{rs} - \delta_{qr}\hat{E}_{ps}
\eea
is the singlet two-electron excitation operator~\cite{pink_bible}. In both Eq.~\eqref{eqn:electronic_H} and Eq.~\eqref{eqn:2e_exc_operator}, the indices $p$, $q$, $r$ and $s$ run over the spatial orbital basis. The \ac{vqe} algorithm \cite{Peruzzo_2014} assumes a pre-defined ansatz structure for $\hat{U}(\bm{\kappa})\hat{U}(\bm{\theta})$ and relies on a quantum-classical strategy to minimize $E(\bm{\kappa},\bm{\theta})$: the quantum computer is used to measure $E(\bm{\kappa},\bm{\theta})$ (and its gradient), while a classical routine is used to optimize either $\bm{\theta}$ alone in the fixed-orbital variants \cite{Peruzzo_2014}, where the orbitals are kept fixed to those of $\ket{\Psi_{\text{ref}}}$, or both $\bm{\theta}$ and $\bm{\kappa}$ in the orbital-optimized variants \cite{Mizukami_2020,Sokolov_2020}.

In the fixed-orbital variants, the \ac{adaptVQE} algorithm considers $\hat{U}(\bm{\kappa})=\hat{I}$ and iteratively constructs a \ac{ups} ansatz~\cite{Evangelista_Chan_Scuseria_2019}
\bea\label{eqn:ups_ansatz}
    \ket{\Psi^{(n)}(\bm{\theta})} = \hat{U}^{(n)}(\theta^{(n)})\dots\hat{U}^{(1)}(\theta^{(1)})\ket{\Psi_{\text{ref}}}.
\eea
Here, 
\bea\label{eqn:unitary_definition}
    \hat{U}^{(n)}(\theta^{(n)})=\exp{(\theta^{(n)}\hat{\tau}^{(n)})},
\eea
where the superscript $n$ refers to the ADAPT iteration and $\hat{\tau}$ represents an anti-Hermitian generator. The ADAPT procedure starts from the definition of a pool $\mathcal{A}=\{\hat{\tau}_g\}$; at each iteration, the algorithm selects a generator from $\mathcal{A}$ and appends the corresponding unitary $\hat{U}^{(n+1)}(\theta^{(n+1)})$ to the circuit, subsequently optimizing
\bea\label{eqn:fo_ADAPT_cost_function}
    E^{(n+1)}(\bm{\theta}) = \bra{\Psi^{(n+1)}(\bm{\theta})}\hat{H}\ket{\Psi^{(n+1)}(\bm{\theta})}
\eea
via a \ac{vqe} procedure. In the fixed-orbital models, we use the universal pool \cite{Evangelista_Chan_Scuseria_2019}
\bea\label{eqn:A_sd_spin}
    \mathcal{A}_{\text{sd}} = \left\{\hat{\tau}_{q\alpha}^{p\alpha},\hat{\tau}_{q\beta}^{p\beta},\hat{\tau}_{q\alpha s\alpha}^{p\alpha r\alpha},\hat{\tau}_{q\beta s\beta}^{p\beta r\beta}, \hat{\tau}_{q\alpha s\beta}^{p\alpha r\beta},\hat{\tau}_{q\beta s\alpha}^{p\beta r\alpha}\right\}.
\eea
Here, the generalized single and double excitations are defined as
\bea\label{eqn:gen_singles}
    \hat{\tau}_{q\sigma}^{p\sigma}=\hat{a}_{p\sigma}^{\dagger}\hat{a}_{q\sigma} - \hat{a}_{q\sigma}^{\dagger}\hat{a}_{p\sigma},
\eea
\bea\label{eqn:gen_doubles}
    \hat{\tau}_{q\sigma s\tau}^{p\sigma r\tau} = \hat{a}_{p\sigma}^{\dagger}\hat{a}_{r\tau}^{\dagger}\hat{a}_{s\tau}\hat{a}_{q\sigma} - \hat{a}_{q\sigma}^{\dagger}\hat{a}_{s\tau}^{\dagger}\hat{a}_{r\tau}\hat{a}_{p\sigma},
\eea
where the spin variables $\sigma,\tau\in\{\alpha,\beta\}$. The $\mathcal{A}_{\text{sd}}$ pool conserves the $\mathcal{S}_z$ and number symmetries, while it does not enforce the $\mathcal{S}^2$ symmetry. In the orbital-optimized models~\cite{OO_ADAPT_VQE_2024}, we consider
\bea\label{eqn:orbital_optimized_op}
    \hat{U}(\bm{\kappa}) = \exp\left\{\sum_{p>q}\kappa_{pq}(\hat{E}_{pq}-\hat{E}_{qp})\right\},
\eea
and, instead of applying $\hat{U}(\bm{\kappa})$ to the wave function, we use it to transform $\hat{H}$ to
\bea\label{eqn:H_orbital_opt}
    \hat{H}(\bm{\kappa}) = \sum_{pq}h_{pq}(\bm{\kappa})\hat{E}_{pq} + \frac{1}{2}\sum_{pqrs}g_{pqrs}(\bm{\kappa})\hat{e}_{pqrs},
\eea
where the effects of the orbital rotations are folded in the Hamiltonian integrals. At each iteration, following the selection step, the energy expression
\bea\label{eqn:oo-En_cost_function}
    E^{(n+1)}(\bm{\kappa},\bm{\theta}) = \bra{\Psi^{(n+1)}(\bm{\theta})}\hat{H}(\bm{\kappa})\ket{\Psi^{(n+1)}(\bm{\theta})}
\eea
is minimized with respect to $\bm{\theta}$ and $\bm{\kappa}$ according to the orbital-optimized \ac{vqe} protocol \cite{Sokolov_2020,Mizukami_2020}. The orbital transformation generates a basis in which the contribution of the single excitations vanishes \cite{pink_bible}. Accordingly, in the orbital-optimized models we exclude the generalized singles excitations from $\mathcal{A}_{\text{sd}}$, restricting it to 
\begin{equation}\label{eqn:pool_oo}
    \mathcal{A}_\textrm{d} = \left\{\hat{\tau}_{q\alpha s\alpha}^{p\alpha r\alpha},\hat{\tau}_{q\beta s\beta}^{p\beta r\beta}, \hat{\tau}_{q\alpha s\beta}^{p\alpha r\beta},\hat{\tau}_{q\beta s\alpha}^{p\beta r\alpha}\right\}.
\end{equation}
The selection of the best operator plays a crucial role in the adaptive algorithms, as the quality of the ansatz depends on the ordering and nature of the operators composing it \cite{Burton_DISCO_2023,Grimsley_Mayhall_2020}. At each iteration, the algorithm establishes a hierarchy of the generators in the pool according to a scoring criterion. The generator associated with the best score is chosen and the corresponding unitary is appended to the \ac{ups} ansatz. In this work, we compare two scoring criteria: gradient-based and energy-based.

\subsubsection{Gradient-based selection}
The gradient-based criterion, introduced by the original fermionic \ac{adaptVQE} algorithm \cite{ADAPT_2019}, is based on the gradient 
\bea\label{eqn:en_gradient}
    \left.
    \frac{\partial E^{(n)}}{\partial 
    \theta_g}\right|_{\theta_g=0} 
    = \bra{\Psi^{(n)}(\bm{\theta})}[\hat{H},\hat{\tau}_g]\ket{\Psi^{(n)}(\bm{\theta})}.
\eea
associated to a given candidate generator $\hat{\tau}_g\in\mathcal{A}$. The selection procedure uses the absolute value of the gradient to score and rank the generators in the pool; the generator associated with the largest gradient score is selected and the corresponding $U^{(n+1)}(\theta^{(n+1)})$ is appended to the \ac{ups} ansatz. The parameter $\theta^{(n+1)}$ is initialized to 0 and brought to its optimal value by the subsequent \ac{vqe} optimization procedure.

\subsubsection{Energy-based selection}
The energy-based criterion scores the generators according to the energy reduction they bring about when the corresponding unitary is appended to the \ac{ups} ansatz. In particular, the score assigned to each candidate generator $\hat{\tau}_g$ corresponds to the minimum of the \ac{1-cf}
\bea\label{eqn:cost_function_energy}
    E_g^{(n+1)}(\theta_g) = \bra{\Psi^{(n)}(\bm{\theta})}e^{-\theta_g\hat{\tau}_g}\hat{H}e^{\theta_g\hat{\tau}_g}\ket{\Psi^{(n)}(\bm{\theta})},
\eea
where the parameters of the previous $n$ layers are kept fixed.

As shown in several examples (including the Rotosolve optimizer \cite{Vidal_2018,Parrish_2019,Nakanishi_2020}, the Rotoselect algorithm \cite{Ostaszewski_2021}, and the generalized parameter-shift rule for quantum gradients \cite{Wierichs_2022}), the \ac{1-cf} can be written as an analytical trigonometric function. For generators satisfying $\hat{\tau}_g^3=-\hat{\tau}_g$ and $\hat{\tau}_g^2\neq\hat{I}$, the unitary operator $e^{\theta_g\hat{\tau}_g}$ can be expressed as (for details, 
see supplementary information)
\bea
\label{eqn:unitary_trigonometric}
    e^{\theta_g\hat{\tau}_g} = \hat{I} + \sin(\theta_g)\hat{\tau}_g + [\cos(\theta_g)-1]\hat{\tau}_g^2.
\eea
This particular expression for $e^{\theta_g\hat{\tau}_g}$ leads, upon insertion in 
Eq.~\eqref{eqn:cost_function_energy} (for details see supplementary information), to an analytical trigonometric expression of the \ac{1-cf}: 
\bea\label{eqn:en_based-landscape}
    E_g^{(n+1)}(\theta_g) = a_{1,g}\cos{(\theta_g)} + b_{1,g}\sin{(\theta_g)} + a_{2,g}\cos{(2\theta_g)} + b_{2,g}\sin{(2\theta_g)} + c_g.
\eea

Determining the explicit form of $E^{(n+1)}_g(\theta_g)$ relies on obtaining the coefficients $a_{1,g}$, $a_{2,g}$, $b_{1,g}$, $b_{2,g}$, $c_g$; this can be done via the evaluation of $E^{(n+1)}_g(\theta_g)$---by measuring the expectation value of $\hat{H}$ in Eq.~\eqref{eqn:cost_function_energy}---at 5 different values of $\theta_g\in[0,2\pi)$ and solving the resulting linear system of equations. By finding the minimum of the \ac{1-cf}, $E^{(n+1)}_g(\theta_g^*)$, one obtains both the energy score for each generator and the optimal parameter $\theta_g^*$. The selection procedure selects the generator corresponding to the lowest energy score and the corresponding unitary operator is appended to the ansatz. The new parameter $\theta^{(n+1)}$ is then initialized to $\theta_g^*$. Compared to the gradient-based approach, where the parameter of the appended operator is initialized to zero, the initialization to $\theta_g^*$ can provide a warm start to the subsequent optimization of the parameters \cite{ExcitationSolve_2025,GGA_ADAPT_2025}. 

The energy-based selection procedure adopted in the current state-of-the-art Rotoselect implementations \cite{Ostaszewski_2021,GGA_ADAPT_2025,ExcitationSolve_2025} further optimizes the reconstruction of the 
\ac{1-cf} by using the energy from the previous selection-optimization iteration, $E^{(n)}$, in correspondence to $\theta_g=0$. In the fermionic variants of Rotoselect \cite{GGA_ADAPT_2025,ExcitationSolve_2025}, 
this enables the reconstruction of
$E^{(n+1)}_g(\theta_g)$ 
from four evaluations of the expectation value of $\hat{H}$ at four distinct nonzero values of $\theta_g\neq 0$.

In the next section, we present our resource-efficient Rotoselect algorithm, which aims at determining $E^{(n+1)}_g(\theta_g)$ with less than four energy evaluations.
   
\subsection{Resource-efficient Rotoselect algorithm}\label{sec:RSe_alg}

\begin{figure}[t!]
    \centering
    \includegraphics[width=0.5\linewidth]{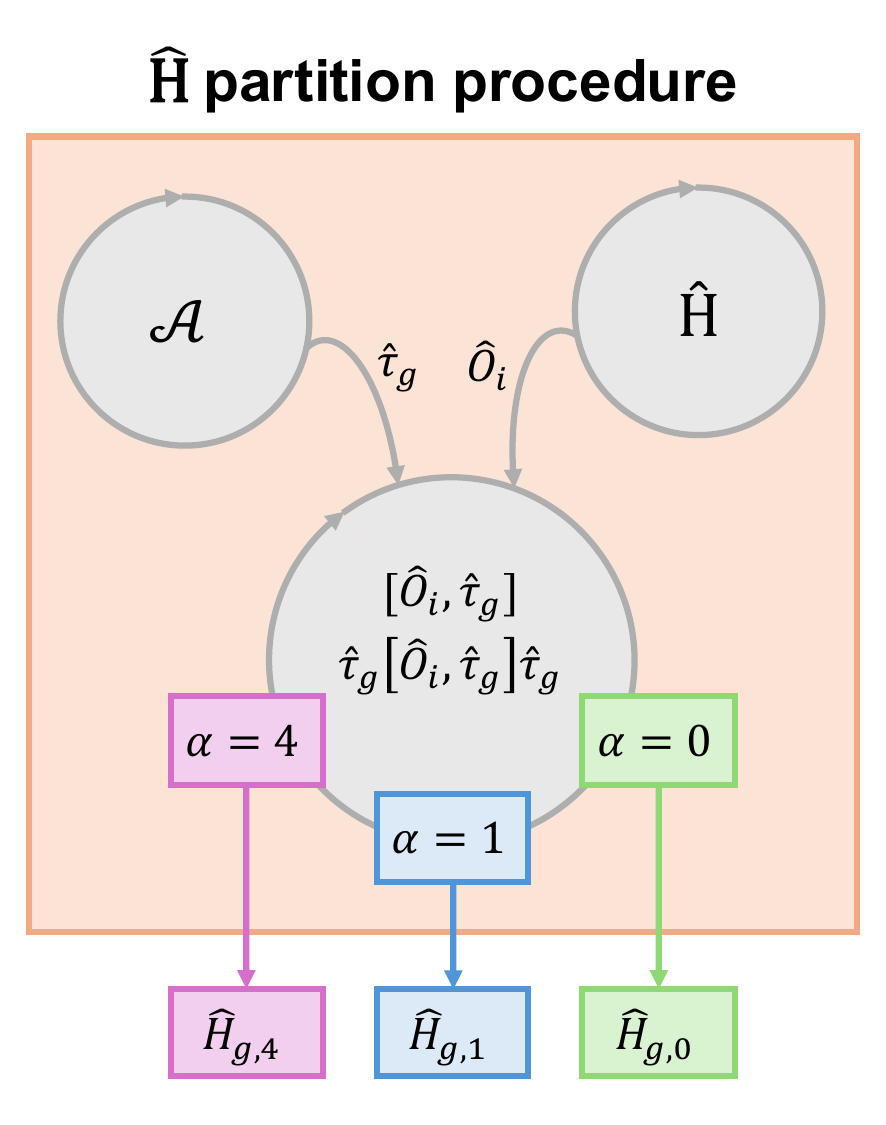}
    \caption{Schematic representation of the generator-dependent procedure for the partitioning of the Hamiltonian. Given a candidate generator $\hat{\tau}_g$ from the pool $\mathcal{A}$, each Fermi string $\hat{O}_i$ from the Hamiltonian is categorized either in the $\alpha=0$, $\alpha=1$ or $\alpha=4$ case according to the value of $[\hat{O}_i,\hat{\tau}_g]$ and $\hat{\tau}_g[\hat{O}_i,\hat{\tau}_g]\hat{\tau}_g$. 
    Following this categorization, each element of the Hamiltonian, $h_i\hat{O}_i$, is assigned to one of the fragments $\hat{H}_{g,0}$, $\hat{H}_{g,1}$ or $\hat{H}_{g,4}$.}
    \label{fig:H_splitting}
\end{figure}

The function $E^{(n+1)}_g(\theta_g)$ in Eq.~\eqref{eqn:cost_function_energy} can also be viewed as the expectation value of a generator-dependent, unitarily transformed Hamiltonian,
\bea\label{eqn:unitary_transf_H}
    \hat{H}_g' = e^{-\theta_g\hat{\tau}_g}\hat{H}e^{\theta_g\hat{\tau}_g}.
\eea
This transformation can be evaluated using a special case of the \ac{bch} expansion \cite{pink_bible},
\bea\label{eqn:bch_expansion}
    \hat{H}'_g = \sum_{k=0}^{\infty}\frac{(-1)^k\theta_g^k}{k!}\text{ad}^k_{\hat{\tau}_g}(\hat{H}),
\eea
where $\text{ad}^k_{\hat{\tau}_g}(\hat{H})$ represents the $k^{th}$-order nested commutator of $\hat{\tau}_g$ and $\hat{H}$. In our application, we define the Hamiltonian $\hat{H}$ in Eq.~\eqref{eqn:electronic_H} as
\bea\label{eqn:electronic_H_Fermi_words}
    \hat{H} = \sum_{i\in\mathcal{K}}h_i\hat{O}_i,
\eea
where the set $\mathcal{K}$ contains all possible combinations of indices $p, q$ and $p, q, r, s$ in Eq.~\eqref{eqn:electronic_H} and each $\hat{O}_i$ corresponds to a product of fermionic creation/annihilation operators (i.e., a Fermi string). As shown in particular by Evangelista and Magoulas \cite{Evangelista_Magoulas_2025} (and similarly in  Refs.~\cite{Izmaylov_Lang_Yen_2021,Kottmann_Anand_Aspuru-Guzik_2021,Jayakumar_Zeng_Izmaylov_2026}), a Fermi string $\hat{O}'_{i,g}$, transformed with respect to an anti-Hermitian fermionic generator $\hat{\tau}_g$ satisfying $\hat{\tau}_g^3=-\hat{\tau}_g$ and $\hat{\tau}_g^2\neq\hat{I}$, can be expanded in a closed form as
\bea\label{eqn:unitary_transf_O}
    \hat{O}_{i,g}' = \hat{O}_i + [\hat{O}_i,\hat{\tau}_g]\frac{\sin{(\sqrt{\alpha}\theta_g)}}{\sqrt{\alpha}} + [[\hat{O}_i,\hat{\tau}_g],\hat{\tau}_g]\frac{1-\cos{(\sqrt{\alpha}\theta_g})}{\alpha}.
\eea
Here, the $\alpha$ parameter refers to the closure relationship
\bea\label{eqn:closure_triple}
    [[[\hat{O}_i,\hat{\tau}_g],\hat{\tau}_g],\hat{\tau}_g] = -[\hat{O}_i,\hat{\tau}_g]-3\hat{\tau}_g[\hat{O}_i,\hat{\tau}_g]\hat{\tau}_g = -\alpha[\hat{O}_i,\hat{\tau}_g],
\eea
where $\alpha$ can assume three values:
\begin{itemize}
    \item $\alpha=1$, when $[\hat{O}_i,\hat{\tau}_g]\neq0$ and $\hat{\tau}_g[\hat{O}_i,\hat{\tau}_g]\hat{\tau}_g=0$;
    \item $\alpha=4$, when $[\hat{O}_i,\hat{\tau}_g]=\hat{\tau}_g[\hat{O}_i,\hat{\tau}_g]\hat{\tau}_g$;
    \item $\alpha=0$, when $[\hat{O}_i,\hat{\tau}_g]=0$ and Eq.~\eqref{eqn:unitary_transf_O} reduces to $\hat{O}_{i,g}'=\hat{O}_i$.
\end{itemize}
For each candidate generator $\hat{\tau}_g$, we categorize the Fermi strings forming $\hat{H}$ based on the $\alpha$ parameter. As illustrated in Fig.~\ref{fig:H_splitting}, this allows us to partition $\hat{H}$ in three fragments, $\hat{H}_{g,0}$, $\hat{H}_{g,1}$, and $\hat{H}_{g,4}$, each containing Fermi strings (multiplied by the corresponding electronic integral) characterized by one of the values of the $\alpha$ parameter.
Because the Fermi strings classification depends on $\hat{\tau}_g$, the partition of $\hat{H}$ is specific to each candidate generator. Based on the generator-dependent partition of $\hat{H}$, the corresponding \ac{1-cf} decomposes into three contributions:
\bea
\label{eqn:cost_function_split}
    E^{(n+1)}_g(\theta_g) = E^{(n+1)}_{g,0} + E^{(n+1)}_{g,1}(\theta_g) + E^{(n+1)}_{g,4}(\theta_g).
\eea
Here, 
\bea
\label{eqn:cost_function_0}
    E_{g,0}^{(n+1)} = \bra{\Psi^{(n)}(\bm{\theta})}\hat{H}_{g,0}\ket{\Psi^{(n)}(\bm{\theta})}
\eea
is a constant corresponding to the $\alpha=0$ case,
\bea\label{eqn:cost_function_1}
    \begin{split}
            E_{g,1}^{(n+1)}(\theta_g) &= \underbrace{\bra{\Psi^{(n)}(\bm{\theta})}\hat{H}_{g,1}\ket{\Psi^{(n)}(\bm{\theta})}}_{c_{g,1}}+\underbrace{\bra{\Psi^{(n)}(\bm{\theta})}[\hat{H}_{g,1},\hat{\tau}_g]\ket{\Psi^{(n)}(\bm{\theta})}}_{a_{g,1}}\sin(\theta_g)\\ 
            &+ \underbrace{\bra{\Psi^{(n)}(\bm{\theta})}[[\hat{H}_{g,1},\hat{\tau}_g],\hat{\tau}_g]\ket{\Psi^{(n)}(\bm{\theta})}}_{b_{g,1}}[1-\cos{(\theta_g)}]
    \end{split}
\eea
corresponds to the $\alpha=1$ case, while
\bea\label{eqn:cost_function_4}
    \begin{split}
        E_{g,4}^{(n+1)}(\theta_g) &= \underbrace{\bra{\Psi^{(n)}(\bm{\theta})}\hat{H}_{g,4}\ket{\Psi^{(n)}(\bm{\theta})}}_{c_{g,4}}+\underbrace{\bra{\Psi^{(n)}(\bm{\theta})}[\hat{H}_{g,4},\hat{\tau}_g]\ket{\Psi^{(n)}(\bm{\theta})}}_{a_{g,4}}\frac{\sin(2\theta_g)}{2}\\ &+\underbrace{\bra{\Psi^{(n)}(\bm{\theta})}[[\hat{H}_{g,4},\hat{\tau}_g],\hat{\tau}_g]\ket{\Psi^{(n)}(\bm{\theta})}}_{b_{g,4}}\frac{1-\cos{(2\theta_g)}}{4}
    \end{split}
\eea
corresponds to the $\alpha=4$ case. 

Instead of explicitly constructing $[\hat{H}_{g,\alpha},\hat{\tau}_g]$ and $[[\hat{H}_{g,\alpha},\hat{\tau}_g],\hat{\tau}_g]$ and measuring their expectation values, we determine the explicit functional form of $E_{g,1}^{(n+1)}(\theta_g)$ and $E_{g,4}^{(n+1)}(\theta_g)$ through a parameter‑shift procedure using three shifts: 
$E_{g,1}^{(n+1)}(\theta_g)$ and $E_{g,4}^{(n+1)}(\theta_g)$ are evaluated at three different values of $\theta_g$, yielding 
a system of linear equations; 
solving the systems for the coefficients $a_{g,1}, b_{g,1}, c_{g,1}$ and $a_{g,4}, b_{g,4}, c_{g,4}$ yields the explicit forms of $E_{g,1}^{(n+1)}(\theta_g)$ and $E_{g,4}^{(n+1)}(\theta_g)$.
The full reconstruction of the \ac{1-cf} is completed by one single evaluation of the constant term $E_{g,0}^{(n+1)}$, which is summed to $E_{g,1}^{(n+1)}(\theta_g)$ and $E_{g,4}^{(n+1)}(\theta_g)$ to give $E_g^{(n+1)}(\theta_g)$. 

We note that, in the fermionic as well as qubit basis, the full Hamiltonian, $\hat{H}$, and the combined fragments, $\{\hat{H}_{g,\alpha}\}$, share the same strings. Thus, the cost of a single full Hamiltonian evaluation is identical to evaluating all three fragments. Owing to the single expectation value of $\hat{H}_{g,0}$ required to obtain $E^{(n+1)}_{g,0}$, the implementation presented so far allows to reconstruct $E^{(n+1)}_g(\theta_g)$ with less than 3 ``full-dimensional'' evaluations of $\hat{H}$. This improves on the implementation presented in \cite{GGA_ADAPT_2025,ExcitationSolve_2025}, which requires four full-dimensional evaluations of $\hat{H}$ to obtain $E^{(n+1)}_g(\theta_g)$. 

Similarly to the implementations in Refs.~\cite{Ostaszewski_2021,GGA_ADAPT_2025,ExcitationSolve_2025}, we further optimize our \ac{1-cf} reconstruction protocol by considering that $E^{(n+1)}_g(0)$ corresponds to $E^{(n)}$.
\begin{figure*}[h]
    \centering
     \includegraphics[width=\textwidth]{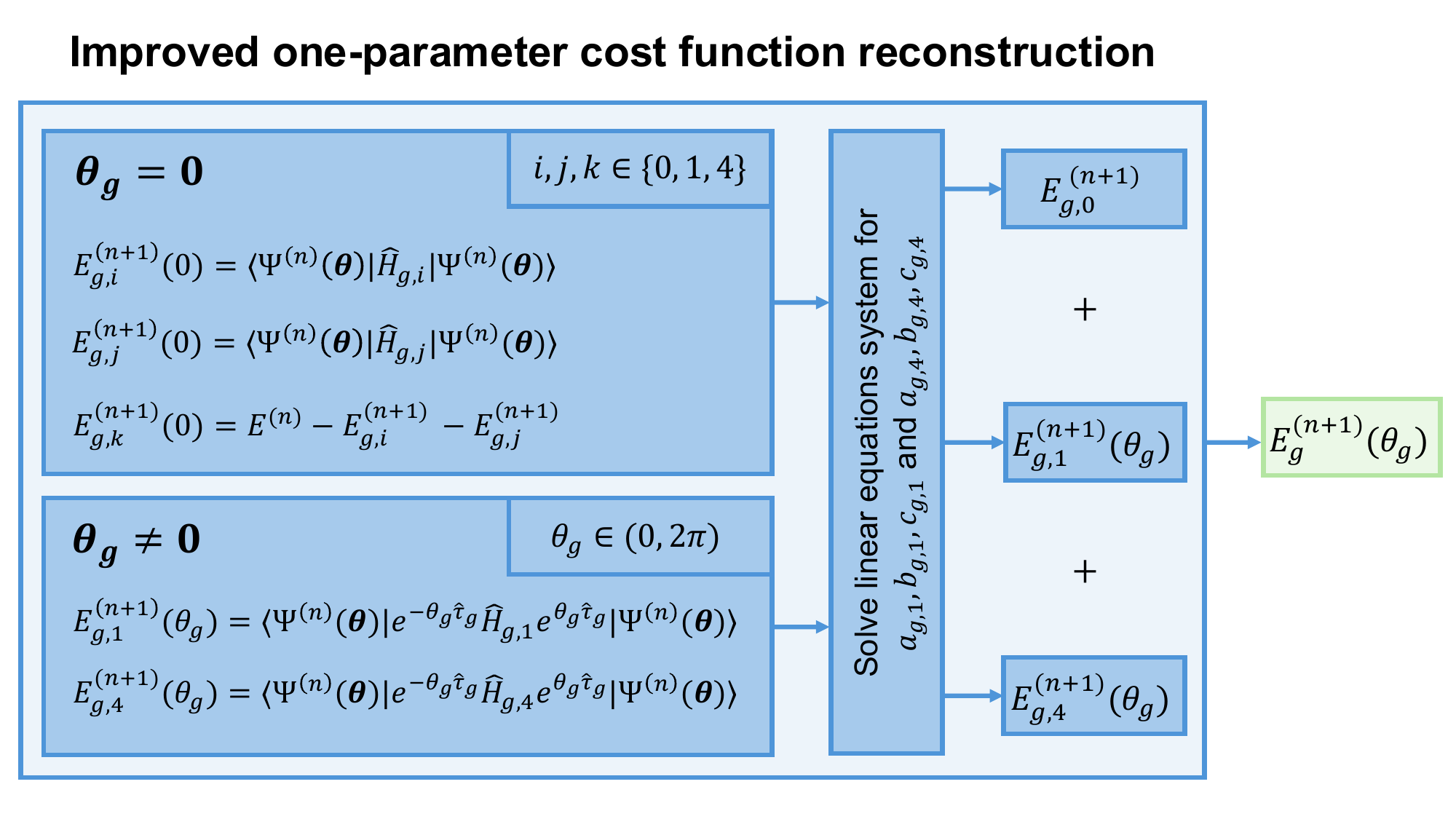}
    \caption{Schematic representation of the resource-efficient procedure to obtain the \acl{1-cf} $E^{(n+1)}_g(\theta)$ associated to each generator $\hat{\tau}_g$. When $\theta_g=0$, we measure the expectation values of the two smallest fragments of the Hamiltonian, $E^{(n+1)}_{g,i}(0)$ and $E^{(n+1)}_{g,j}(0)$. We obtain the third expectation value, $E^{(n+1)}_{g,k}(0)$, by subtracting $E^{(n+1)}_{g,i}(0)$ and $E^{(n+1)}_{g,j}(0)$ from the optimized energy of the previous iteration, $E^{(n)}$. When $\theta_g\neq0$, we only measure $E^{(n+1)}_{g,1}(\theta_g)$ and $E^{(n+1)}_{g,4}(\theta_g)$. We solve the systems of equations to obtain $E^{(n+1)}_{g,1}(\theta_g)$ and $E^{(n+1)}_{g,4}(\theta_g)$, which summed to $E^{(n+1)}_{g,0}$ yield $E^{(n+1)}_g(\theta_g)$.} 
    \label{fig:efficient_cost_function}
\end{figure*}
However, unlike in the original Rotoselect implementations, we cannot simply set $E^{(n+1)}_g(0)$ equal to $E^{(n)}$. As shown in Eq.~\eqref{eqn:cost_function_split}, $E^{(n+1)}_g(0)$ is decomposed in three components, with a pattern characteristic of each generator in the pool. Thus, in order to reconstruct $E^{(n+1)}_g(0)$, we still need the values of $E^{(n+1)}_{g,1}(0)$, $E_{g,4}^{(n+1)}(0)$ and $E_{g,0}^{(n+1)}$ characteristic of each candidate generator.
Despite this, to reduce the cost of evaluating $E^{(n+1)}_g(0)$, we can exploit the fact that the sum of $E^{(n+1)}_{g,1}(0)$, $E_{g,4}^{(n+1)}(0)$ and $E_{g,0}^{(n+1)}$ equals $E^{(n)}$. Accordingly, as shown in the $\theta_g=0$ panel of Fig.~\ref{fig:efficient_cost_function}, to obtain $E_{g,1}^{(n+1)}(0)$, $E_{g,4}^{(n+1)}(0)$ and $E_{g,0}^{(n+1)}$ one needs to measure the expectation values of only two Hamiltonian fragments; the third expectation value can then be obtained by subtracting the measured expectation values from $E^{(n)}$. By measuring only the expectation values of the two ``smallest'' Hamiltonian fragments (i.e., those associated with the lowest number of Pauli strings), a maximal cost reduction is obtained.

\begin{figure*}[t!]
    \centering
    \includegraphics[width=\textwidth]{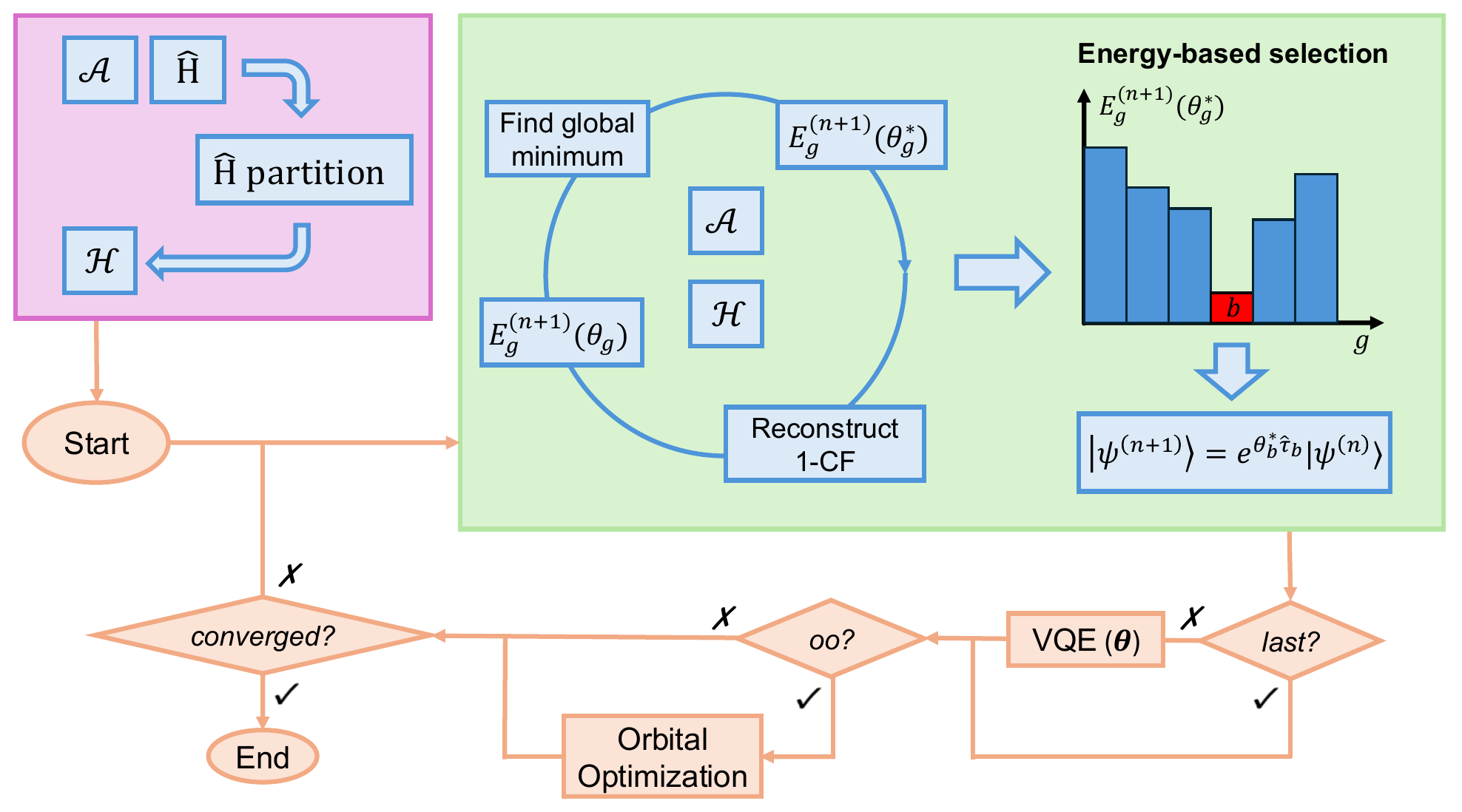}
    \caption{Flowchart of the adaptive algorithm implementing the resource-efficient Rotoselect selection routine. Before starting the ansatz construction, the algorithm partitions the Hamiltonian in fragments characteristic of each generator in $\mathcal{A}$. The pool of Hamiltonian fragments $\mathcal{H}$ is created. The algorithm then enters the adaptive protocol, wherein, at each layer, the \ac{1-cf} for each candidate generator $\hat{\tau}_g$ in the pool is determined using our resource-efficient cost function method. Similarly to Ref.~\cite{ExcitationSolve_2025}, we obtain the  minimum of the \ac{1-cf} associated to each generator (i.e., the energy score) and the corresponding $\theta^*_g$ parameter using the companion matrix method. The best generator, $\hat{\tau}_b$, is chosen according to its energy score and the respective unitary is appended to the ansatz with its parameter initialized to $\theta^*_b$. The ansatz is then re-optimized according to a variety of approaches, involving both the parameters and the orbitals.}
    \label{fig:RSe-scheme}
\end{figure*}

We incorporate the generator-dependent Hamiltonian partition routine and improved \ac{1-cf} reconstruction in the algorithm workflow represented in Fig.~\ref{fig:RSe-scheme}. 
In practice, the Hamiltonian partition routine is performed once before the adaptive ansatz construction. This yields a pool $\mathcal{H}$ of Hamiltonian fragments, where each set of fragments uniquely corresponds to a given generator $\hat{\tau}_g \in \mathcal{A}$. For each generator in $\mathcal{A}$, the corresponding set of Hamiltonian fragments from $\mathcal{H}$ is used to determine the energy score for each generator. 

After the (either gradient-based or energy-based) selection of a new operator, we optimize $\ket{\Psi^{(n+1)}(\bm{\theta},\bm{\kappa})}$ in four different ways, combining different strategies for the $\bm{\theta}$ and $\bm{\kappa}$ optimization. 
We optimize $\bm{\theta}$ following two strategies: \textit{last}, where we either assign $\theta^{(n+1)}$ to the optimal value output by the RSe selection procedure (as shown in Fig.~\ref{fig:RSe-scheme}) or, when the gradient-based selection strategy is adopted, we run a \ac{vqe} optimization of only $\theta^{(n+1)}$; \textit{full}, where we run a \ac{vqe} optimization including all parameters in $\bm{\theta}$. 
Two strategies also adopted in relation to orbital optimization: \textit{fixed-orbital}, where the orbitals remain unchanged and $\mathcal{A}=\mathcal{A}_{\text{sd}}$; \textit{orbital-optimized}, where we optimize $\bm{\kappa}$ at each iteration, considering $\mathcal{A}=\mathcal{A}_\textrm{d}$. The orbital-optimized strategy avoids the quantum measurements associated with selecting single excitations, folding their effects into the electronic integrals of the Hamiltonian via an entirely classical orbital optimization.

\clearpage

\section{Computational details}
We performed the implementation and the calculations at the state vector level using the \textsc{SlowQuant} package \cite{SlowQuant}, interfaced with the \textsc{PySCF} quantum chemistry package \cite{pySCF} for the electronic integrals composing the Hamiltonian.  
The notation ($n_e$, $n_o$) is used throughout to specify the number of electrons $n_e$ and number of spatial orbitals $n_o$ for each system. All calculations consider the complete active space spanned by the STO-3G basis.
For the three benchmark molecules 
at their equilibrium geometries, we adopted the following bond lengths: 
a Li-H bond length of 
1.595 {\AA} for LiH; 
a Be-H bond length of 
1.334 {\AA} for linear \ce{BeH2}; 
a O-H bond length of 0.96 {\AA} for \ce{H2O} bent at 104.5$^{\circ}$. 
At the stretched geometries, we used the following bond lengths:
3 {\AA} for LiH; 2.668 {\AA} for the symmetrically linearly stretched \ce{BeH2}; 1.81 {\AA} for the symmetrically stretched, bent \ce{H2O}. 

To ensure a fair comparison between the selection strategies, we set the maximum number of ansatz parameters to one hundred and utilized the same Sequential Least Squares Programming (SLSQP) optimizer, as implemented in the \textsc{SciPy} package \cite{2020SciPy-NMeth}, for all the algorithms 
including parameter re-optimization. 

We assess the measurement cost of each method using a noiseless circuit-evaluation proxy.
Specifically, we evaluate the cost of a single energy or gradient evaluation according to the number of Pauli strings appearing in the Jordan-Wigner decomposition~\cite{Jordan_Wigner_1928} of the measured operator. For energy-based selection, this corresponds to the full Hamiltonian or the corresponding fragments; for gradient-based selection, this corresponds to the commutator operator in Eq.~\eqref{eqn:en_gradient}. We obtain the total cost of an adaptive run by summing the total number of Pauli strings measured during the selection and optimization steps. To establish a uniform comparison between different strategies for adaptive ansatz construction (in particular between fermionic Rotoselect and our resource-efficient variant), we consider all Pauli strings without employing any qubit-wise commuting or more general, cross-pool operator-grouping techniques \cite{Anastasiou_2024,Anastasiou_Mayhall_2023}. In particular, we do not include finite-shot noise, measurement-allocation effects, or hardware-dependent compilation overhead. The reported costs should
therefore be interpreted as Pauli-string-weighted counts of noiseless circuit evaluations.

\clearpage
\section{Results}\label{sec:results}
We first assess the cost of the operator-selection step, comparing three fermionic adaptive algorithms:
gradient-based \ac{adaptVQE} (GB) \cite{ADAPT_2019}, 
standard fermionic Rotoselect (RS) \cite{GGA_ADAPT_2025,ExcitationSolve_2025}, and our own resource-efficient Rotoselect implementation (RSe).
\begin{table*}[h]
    \centering
    \setlength{\tabcolsep}{8pt}
    \caption{Ratios of selection cost among RS, RSe, and GB in equilibrium and stretched \ce{LiH}~$(4,6)$, \ce{BeH2}~$(6,7)$, and \ce{H2O}~$(10,7)$. The selection cost of each model is calculated relative to the $\mathcal{A}_{\text{sd}}$ pool associated with each molecular system.}
    \label{tab:model_cost_comparison}
    \begin{tabular}{l ccc ccc}
        \toprule
        & \multicolumn{3}{c}{Equilibrium} & \multicolumn{3}{c}{Stretched} \\
        \cmidrule(lr){2-4} \cmidrule(lr){5-7}
        ~ & \ce{LiH}~$(4,6)$ & \ce{BeH2}~$(6,7)$ & \ce{H2O}~$(10,7)$ & \ce{LiH}~$(4,6)$ & \ce{BeH2}~$(6,7)$ & \ce{H2O}~$(10,7)$ \\
        \midrule
        RS/GB  & 2.13 & 2.18 & 2.48 & 2.13 & 2.18 & 2.08 \\
        RS/RSe & 1.97 & 2.06 & 2.11 & 1.97 & 2.06 & 2.04 \\
        RSe/GB & 1.09 & 1.06 & 1.17 & 1.09 & 1.06 & 1.02 \\
        \bottomrule
    \end{tabular}
\end{table*}

In Table~\ref{tab:model_cost_comparison}, we compare the three algorithms according to the ratio of their cost relative to the $\mathcal{A}_{\text{sd}}$ pool. For all the tested systems, GB is more than two times cheaper than RS. This result differs from that obtained in Ref.~\cite{ExcitationSolve_2025}, where equal costs are reported. This is because the authors used a less measurement-efficient gradient evaluation metric. Instead of a Pauli string decomposition of the commonly used gradient commutator (see Eq.~\eqref{eqn:en_gradient}), they quantified the cost in terms of energy measurements via a 4-shift parameter-shift rule. The RS and RSe strategies are mathematically equivalent and construct identical ans{\"a}tze (as shown in detail in the Supplementary Information). However, RSe reduces the cost for reconstructing the \ac{1-cf} by roughly a factor of two compared to RS. Furthermore, within the context of our ungrouped cost-evaluation convention, RSe reduces the measurement costs to the level of GB's electronic gradient evaluation.

\subsection{Low correlation regime: equilibrium geometries}
\begin{figure*}[t]
    \includegraphics[width=\textwidth]{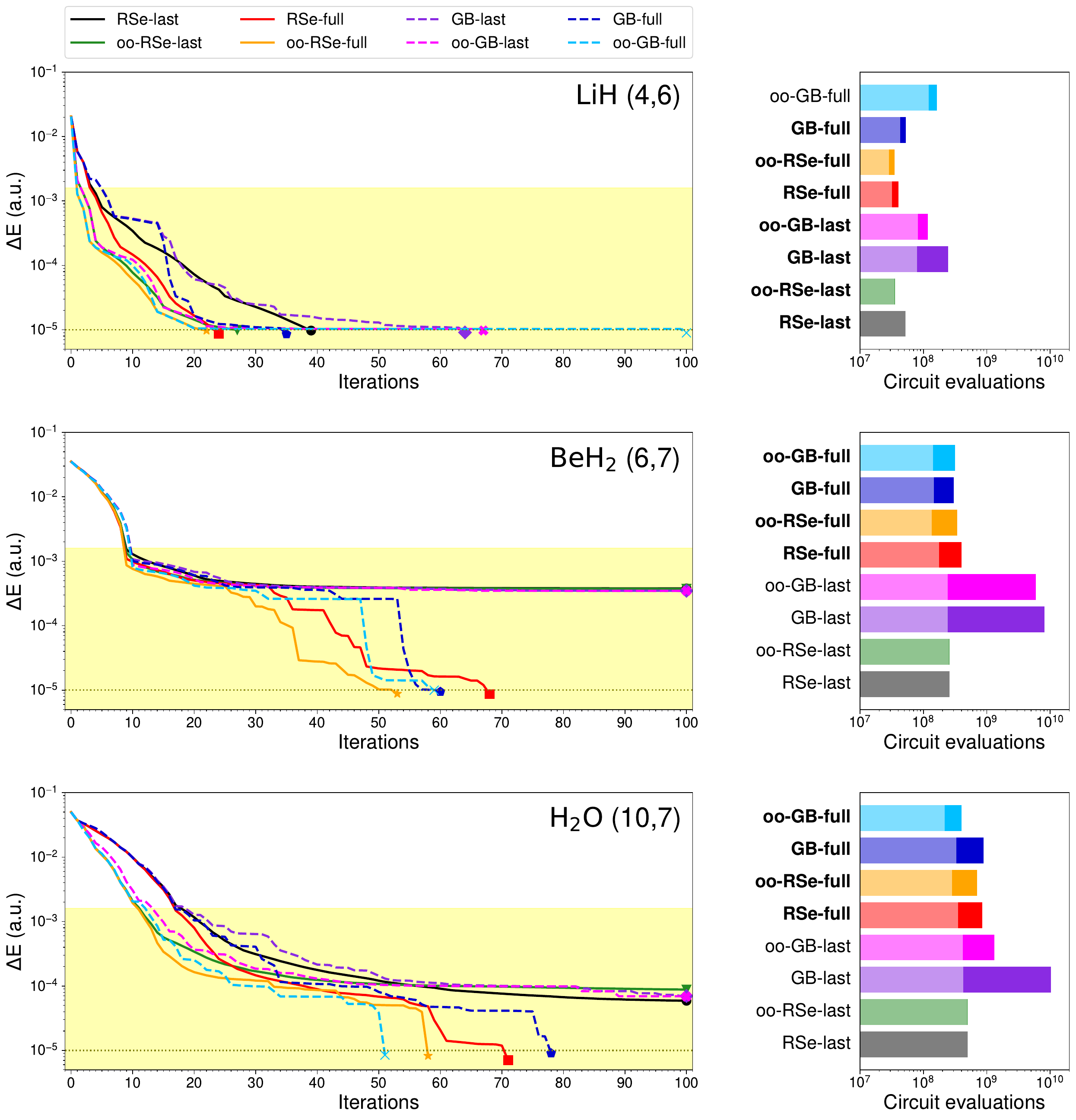}
    \caption{Comparison of selection-optimization strategies in equilibrium LiH (4,6), \ce{BeH2} (6,7) and \ce{H2O} (10,7). Left column: energy convergence with the number of iterations. The light-yellow area indicates the chemical accuracy window relative to the \ac{fci} reference, while the black dotted line represents the convergence threshold ($10^{-5}$ $E_\textrm{h}$ difference). Right column: relative cost in terms of number of circuit evaluations. The lighter/darker part of the bars corresponds to the selection/optimization cost. The models that converged within one hundred iterations are indicated by bold labels.}
    \label{fig:benchmark-eq_geom}
\end{figure*}

We compare the equilibrium geometry calculations in Fig.~\ref{fig:benchmark-eq_geom}. 
In the left panels, we show the energy convergence with the number of iterations; in the right panels, we show the cost in terms of number of noiseless circuit evaluations, separated into selection and optimization contributions.

For LiH~$(4,6)$, all models reach the convergence threshold (i.e., $10^{-5}$ $E_\textrm{h}$ from the \ac{fci} reference) within one hundred iterations. The difference between the selection strategies emerges particularly from the comparison of the \textit{last} models. In both the fixed-orbital and orbital-optimized variants, energy-based selection leads to faster convergence than gradient-based selection, which directly translates into a lower total cost for RSe-based models. In particular, in such low-correlation system, the parameters obtained from the RSe procedure are already close to the optimal value yielded by the subsequent variational optimization.

While full optimization is not needed to reach the convergence threshold, it leads to fewer iterations than the corresponding \textit{last} variants. At the same time, orbital optimization further improves the early-stage energy convergence and reduces the total cost. An exception to this trend is the oo-GB-full model (light-blue, dashed), whose optimization, after an initial swift convergence, stagnates for several layers in the proximity of the convergence threshold. This behavior is consistent with the qualitative onset of a gradient-trough regime~\cite{Grimsley_2023}. 

The best balance between ansatz depth and overall cost for LiH~$(4,6)$ at equilibrium is obtained with oo-RSe-last (green, solid). This model combines the \ac{1-cf}-aware properties of energy-based selection with the faster early convergence of the orbital-optimized formulation, which also leads to a lower selection cost due to the reduced pool size.

The above considerations do not apply to \ce{BeH2}~$(6,7)$ and \ce{H2O}~$(10,7)$. In these systems, the dominant difference is no longer between GB and RSe, but rather between the \textit{last} and \textit{full} re-optimization strategies. In fact, after the first few iterations to reach chemical accuracy, the energy of the \textit{last} models does not improve significantly with the addition of new operators. By contrast, the models adopting full parametric re-optimization manage to evade the stagnating iterations and converge to the set threshold. Thus, with the increase in correlation, the parameter optimization strategy becomes a key determinant of performance. The selection strategy still influences the convergence path, as confirmed by the different optimization paths followed by the RSe and GB models; this seems to stem primarily from the \ac{1-cf}-aware nature of energy-based selection, which leads to the selection of different operators compared to GB. 

In the Supplementary Information, we also quantify the cost saving from RSe's ``warm'' parameter initialization. For last optimization this saves up to $10^5$ circuit evaluations per ADAPT iteration, while for full optimization the cost saving progressively fades with the growth of the \ac{ups} ansatz.

In \ce{BeH2}~$(6,7)$ and \ce{H2O}~$(10,7)$, the effect of orbital-optimization is also clearer than that of the selection strategy. In both systems, the orbital-optimized models lower the energy more effectively in the early ADAPT iterations compared to their fixed-orbital counterparts. As the optimization approaches the convergence threshold, both the orbital-optimized and fixed-orbital models go through stagnation phases. Despite this, the orbital-optimized models retain the initial advantage, reaching the convergence threshold in a lower number of iterations than their fixed-orbital counterparts.

\clearpage
\subsection{Higher correlation regime: stretched geometries}
\begin{figure*}[t]
    \centering
    \includegraphics[width=\textwidth]{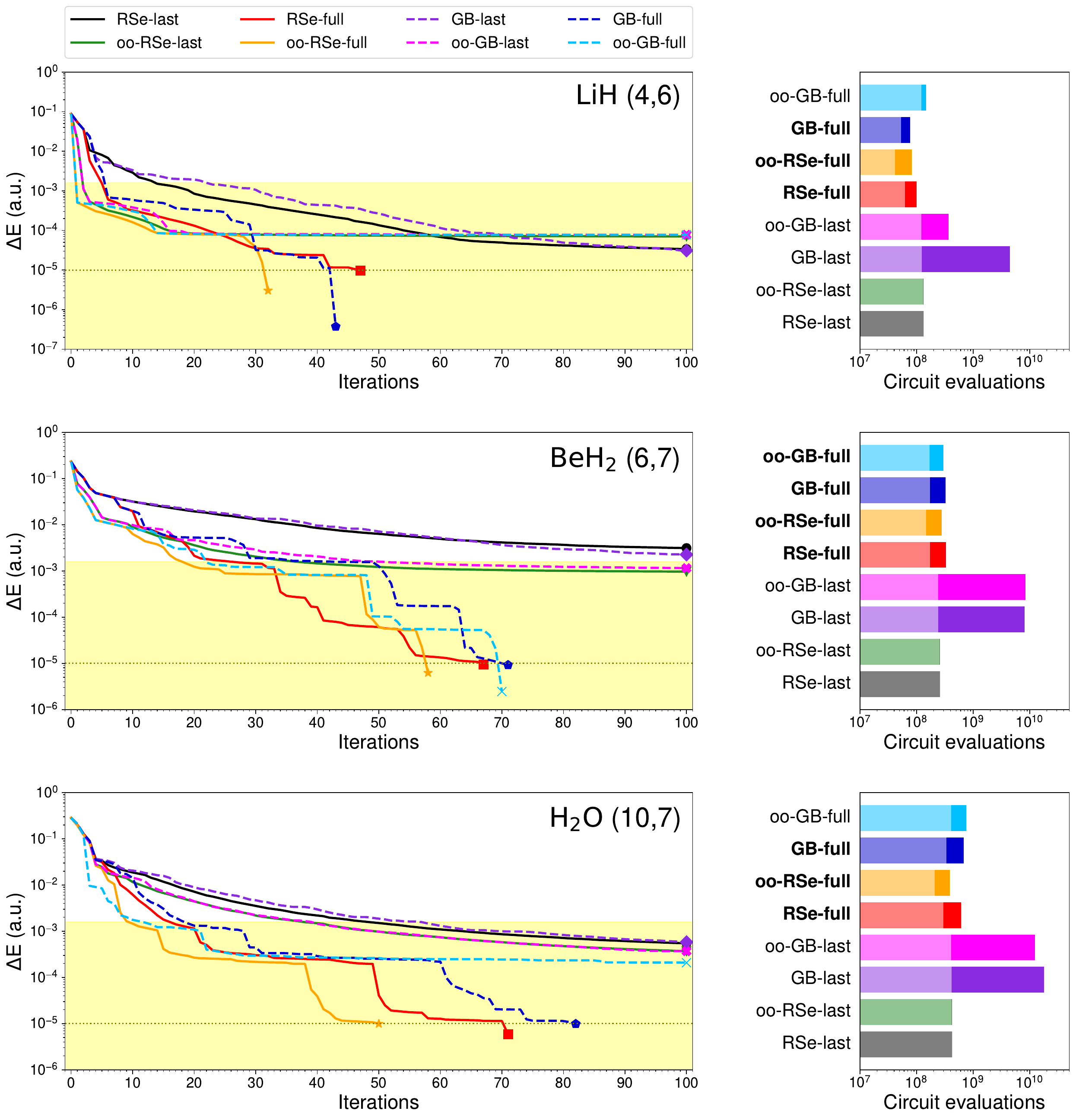}
    \caption{Comparison of selection-optimization strategies in stretched LiH $(4,6)$, \ce{BeH2} $(6,7)$ and \ce{H2O} $(10,7)$. Left column: energy convergence with the number of iterations. Right column: relative cost in terms of number of circuit evaluations. The lighter/darker part of the bars corresponds to the selection/optimization cost. The models that converged within one hundred iterations are indicated by bold labels.}
    \label{fig:benchmark-str_geom}
\end{figure*}
We compare the stretched geometry calculations in \fig{fig:benchmark-str_geom}. By stretching the bonds, we compare the performance of the models at a higher level of correlation. In this regime, the distinction between \textit{last} and \textit{full} optimization becomes even more pronounced.

For stretched LiH~$(4,6)$, the need to describe a more correlated system leads to the same separation between the \textit{last} and \textit{full} models observed in equilibrium \ce{BeH2}~$(6,7)$ and \ce{H2O}~$(10,7)$. Only the models that include a full re-optimization of the ansatz converge to the set threshold. Within the \textit{full} re-optimization models, the comparison between RSe and GB is more favorable to the energy-based strategy than in the equilibrium case. In fact, the RSe-full models consistently reach the target threshold, whereas oo-GB-full (cyan, dashed) again exhibits a marked tendency towards stagnation. At the same time, the comparison between oo-RSe-full (orange, solid) and RSe-full (red, solid) shows how, similarly to the equilibrium geometry case, the orbital-optimization gives an early convergence advantage to the corresponding models. Although both models undergo periods of slow progress, the orbital-optimized variant preserves its initial lead until the convergence to the set threshold, determining shallower circuits and a lower total cost.

The same overall picture is observed for stretched \ce{BeH2}~$(6,7)$ and \ce{H2O}~$(10,7)$: pairing full ansatz re-optimization with energy-based selection and orbital optimization leads to the best overall performance. For the \textit{last} models, the selection strategy does not have a significant impact. On the other hand, orbital optimization plays a distinctive role: in stretched \ce{BeH2}~$(6,7)$, for example, it makes the difference between reaching or not reaching chemical accuracy.

\clearpage
\section{Conclusions}
\label{sec:conclusions}
In this paper, we introduced a lower-cost reconstruction of the fermionic Rotoselect one-parameter cost function (1-CF) based on exact, generator-dependent Hamiltonian transformations. We compared our resource-efficient algorithm to the state-of-the-art fermionic Rotoselect implementations \cite{GGA_ADAPT_2025,ExcitationSolve_2025} and the gradient-based approach as implemented in \ac{adaptVQE} \cite{ADAPT_2019}. Compared to standard fermionic Rotoselect, our resource-efficient implementation delivers the same energy score using, on average, half the number of circuit evaluations (based on an ungrouped Pauli string measurement decomposition). This cost reduction afforded by our resource-efficient implementation allows to deliver the \ac{1-cf}-aware features of the energy-based approach at about the same cost as the gradient-based strategy. We note, however, that further cost-reduction strategies, such as cross-pool grouping of commuting Pauli strings in the gradient-based approach \cite{Anastasiou_Mayhall_2023}, are not reflected in our cost metric. 

We investigated the role of the selection approach relative to other algorithmic choices by performing a benchmark study; specifically, we combined the gradient-based and energy-based selection with a variety of optimization approaches. These included \textit{last} and \textit{full} strategies to parametric optimization, combined with either fixed- or orbital-optimized solutions. As benchmark systems, we considered LiH~$(4,6)$, \ce{BeH2}~$(6,7)$, and \ce{H2O}~$(10,7)$ at their equilibrium and stretched geometries.

The benchmark at equilibrium geometries indicates that the \ac{1-cf}-aware properties afforded by the energy-based strategy are most valuable in the most weakly correlated case (i.e., LiH~$(4,6)$), where the oo-RSe-last model represents a cost-efficient solution that avoids the \ac{vqe} optimization. This could be advantageous in an application on NISQ devices, making oo-RSe-last a promising testbed for future studies under the finite-shot and noisy conditions of current hardware.

With the increase in correlation, the optimization often stagnates as it gets trapped in plateaus of the optimization landscape. In these cases, 
ansatz re-optimization and orbital optimization have a greater impact on the convergence behavior than the selection strategy.
Full ansatz re-optimization becomes crucial to systematically reach convergence, while orbital optimization provides an early-stage advantage, which speeds up convergence and lowers the costs. Within the scope of this benchmark study, the oo-RSe-full model yields the most consistently favorable balance of convergence reliability and lowest total cost.

Future work will investigate whether analogous algebraic simplifications can be derived for other pools such as \ac{qeb} \cite{Yordanov_2021} and \ac{ceo} \cite{Ramoa_2025}, whose generator algebras differ from the fermionic excitation generators considered here. Measurement grouping and other cost-reduction techniques \cite{Ramôa_Santos_Economou_2025,Anastasiou_Mayhall_2023} will also be the subject of future investigations. Additional improvements will focus on alleviating stagnating periods in the adaptive ansatz optimization by enforcing spin and spatial symmetries \cite{Bertels_2022,Tsuchimochi_2022,Magoulas_Zhang_Evangelista_2025} or adopting global optimization strategies \cite{Burton_DISCO_2023,Stadelmann_2025}.

\section*{Conflicts of interest}
There are no conflicts to declare.

\section*{Data availability}
All data supporting the findings of this study are included in the article and the supplementary information. The figures can be reproduced using the datasets and code available at \url{https://doi.org/10.5281/zenodo.21264507}. The algorithm code is available at \url{https://github.com/emarossi/rotoadapt.git}.

\section*{Acknowledgements}
We thank Dr. Anurag Singh (SDU) for providing us with his fermionic gradient-based ADAPT‑VQE implementation in SlowQuant.
We acknowledge useful discussions with Dr. Dipanjali Halder and Theo Juncker Buchwald (DTU).
E.R., S.C. and S.P.A.S acknowledge funding from Innovation Fund Denmark within the Eureka project \textit{Q-Chemion}, 
project numbers 4340-00005B and 4340-00006B.
S.C., K.M.Z., E.R.K, and S.P.A.S. also acknowledge  funding from the Novo Nordisk
Foundation for the focused research project \textit{Hybrid Quantum
Chemistry on Hybrid Quantum Computers}, (HQC)$^2$,
grant
number NNFSA220080996.
K.M.Z. acknowledges funding from the Royal Society of Chemistry Collaboration grant, C25-1492721325.
The Otto M{\o}nsted Fond is acknowledged for a guest professorship at DTU to Prof. Artur Izmaylov (Journal nr. 25-12-2099).

\newpage

\bibliography{references}

\end{document}


\newcommand{\bea}{\begin{eqnarray}}
\newcommand{\eea}{\end{eqnarray}}
\newcommand{\EQ}[1]{Equation~(\ref{#1})} %
\newcommand{\eq}[1]{Eq.~(\ref{#1})} %
\newcommand{\eqs}[1]{Eqs.~(\ref{#1})} %
\newcommand{\fig}[1]{Fig.~\ref{#1}} %
\newcommand{\figs}[1]{Figs.~\ref{#1}} %
\newcommand{\CR}[1]{\hat a^{\dagger}_{#1}}
\newcommand{\AN}[1]{\hat a_{#1}}

\newcommand{\LA}[1]{\mathfrak{#1}}
\newcommand{\HG}{\hat G}

\newcommand{\BC}[1]{\hat \beta^{\dagger}_{#1}}
\newcommand{\BA}[1]{\hat \beta_{#1}}
\newcommand{\EU}[1]{\hat E^{#1}}
\newcommand{\ED}[1]{\hat E_{#1}}
\newcommand{\EE}[2]{\hat E_{#2}^{#1}}
\newcommand{\EX}[2]{\mathcal{E}_{#2}^{#1}}
\newcommand{\GP}[1]{\hat \gamma_{#1}}
\newcommand{\kh}{\hat \kappa}
\newcommand{\OO}{\hat O}
\newcommand{\HC}{\hat C}
\newcommand{\HA}[1]{\hat A_{#1}}


\newcommand{\bfr}{\mathbf{r}} %
\newcommand{\E}{\textrm{e}} %
\newcommand{\I}{\mathrm{i}\mkern1mu} %
\newcommand{\RR}{\mathbf{R}} %
\newcommand{\HP}{\hat P} %
\newcommand{\rr}{\mathbf{r}} %
\newenvironment{NB}{\color{red}NB: }{\ignorespacesafterend}  %
\newcommand{\HH}{\hat H} %
\newcommand{\HU}{\hat U} %
\newcommand{\hz}{\hat z} %

\newcommand{\AFI}[1]{\textcolor{blue}{[AFI: #1]}}
\newenvironment{KMZ}{\par\begingroup\color{orange}}{\par\endgroup}

\title{Supplemental Material for:\\
Resource-efficient energy-based operator selection in fermionic ADAPT-VQE via exact Hamiltonian transformation}

\author{Emanuele Rossi}
\email{emaro@kemi.dtu.dk}
\affiliation{DTU Chemistry, Technical University of Denmark, Kemitorvet 207, 2800 Kongens Lyngby, Denmark}
\author{Erik Rosendahl Kjellgren}
\affiliation{Department of Physics, Chemistry and Pharmacy, University of Southern Denmark, Campusvej~55, DK--5230 Odense M, Denmark}
\author{Artur F. Izmaylov}
\affiliation{Department of Physical and Environmental Sciences,
University of Toronto Scarborough, Toronto, Ontario M1C 1A4, Canada}
\affiliation{Chemical Physics Theory Group, Department of Chemistry, University of Toronto, Toronto, Ontario M5S 3H6, Canada}
\author{Stephan P.A. Sauer}
\affiliation{Department of Chemistry, University of Copenhagen, Universitetsparken 5, 2100 Copenhagen,
Denmark}
\author{Karl Michael Ziems}
\email{K.M.Ziems@soton.ac.uk}
\affiliation{School of Chemistry, University of Southampton, Highfield, Southampton SO17 1BJ, United Kingdom}
\affiliation{DTU Chemistry, Technical University of Denmark, Kemitorvet 207, 2800 Kongens Lyngby, Denmark}
\author{Sonia Coriani}
\affiliation{DTU Chemistry, Technical University of Denmark, Kemitorvet 207, 2800 Kongens Lyngby, Denmark}

\maketitle

\section{Expansion of unitary operator for fermionic generators}\label{app:unitary_expansion}
The one-parameter unitary operator can be Taylor expanded as 
\bea\label{eqn:unitary_expansion_1}
    \hat{U}(\theta_g) = e^{\theta_g\hat{\tau}_g} = \sum_{k=0}^{\infty}\frac{(\theta_g\hat{\tau}_g)^k}{k!}~,
\eea
where $\hat{\tau}_g$ is a generic anti-Hermitian fermionic generator. 
The generator satisfies the properties $\hat{\tau}_g^2\neq\hat{I}$ and 
$\hat{\tau}_g^3=-\hat{\tau}_g$, which reduces every power of $\hat{\tau}_g$ in eq. \eqref{eqn:unitary_expansion_1} to a linear combination of $\{\hat{I}$, $\hat{\tau}_g$, $\hat{\tau}_g^2\}$. 
For example, $\hat{\tau}_g^4$ = $\hat{\tau}_g^3\hat{\tau}_g$ = $-\hat{\tau}_g^2$ or $\hat{\tau}_g^5$ = $\hat{\tau}_g^3\hat{\tau}_g^2$ = $-\hat{\tau}_g^3$ = $\hat{\tau}_g$. This establishes a recursive rule within the Taylor expansion, according to which the odd-power terms can be written as
\bea\label{eqn:unitary_expansion_2}
    \sum_{k=0}^{\infty}\frac{\theta_g^{2k+1}}{(2k+1)!}(-1)^k\hat{\tau}_g = \sin(\theta_g)\hat{\tau}_g~,
\eea
while the even-power terms---apart from the $k=0$ term---can be written as
\bea
\label{eqn:unitary_expansion_3}
    \sum_{k=1}^{\infty}\frac{\theta_g^{2k}}{(2k)!}(-1)^k\hat{\tau}_g^2 = \left(\sum_{k=0}^{\infty}\frac{\theta_g^{2k}}{(2k)!}(-1)^k - 1\right)\hat{\tau}_g^2 = (\cos(\theta_g)-1)\hat{\tau}_g^2~.
\eea
Given the expansions in 
eq.~\eqref{eqn:unitary_expansion_2} and 
eq.~\eqref{eqn:unitary_expansion_3}, we can write
\bea
\label{eqn:unitary_expansion_4}
    \hat{U}(\theta_g) = e^{\theta_g\hat{\tau}_g} = \hat{I} + \sin(\theta_g)\hat{\tau}_g + (\cos(\theta_g)-1)\hat{\tau}_g^2,
\eea
where $\hat{I}$ corresponds to the $k=0$ term in 
eq.~\eqref{eqn:unitary_expansion_1}.

\section{Derivation of 5-term energy cost function}\label{app:4-shifts_cost_function}
By substituting the analytical expression
\bea\label{eqn:unitary_trigonometric}
    e^{\theta_g\hat{\tau}_g} = \hat{I} + \sin(\theta_g)\hat{\tau}_g + (\cos(\theta_g)-1)\hat{\tau}_g^2~.
\eea
into the expression for the energy cost function,
\bea\label{eqn:cost_function_energy}
    E_g^{(n+1)}(\theta_g) = \bra{\Psi^{(n)}(\bm{\theta})}e^{-\theta_g\hat{\tau}_g}\hat{H}e^{\theta_g\hat{\tau}_g}\ket{\Psi^{(n)}(\bm{\theta})},
\eea
developing the products, and reordering the terms we obtain
\bea
    \begin{split}
        E_g^{(n+1)}(\theta_g) &= \bra{\Psi^{(n)}(\bm{\theta})}\hat{H}\ket{\Psi^{(n)}(\bm{\theta})} - \bra{\Psi^{(n)}(\bm{\theta})}\{\hat{H},\hat{\tau}_g^2\}\ket{\Psi^{(n)}(\bm{\theta})}\\
        &+\bra{\Psi^{(n)}(\bm{\theta})}\hat{\tau}_g^2\hat{H}\hat{\tau}_g^2\ket{\Psi^{(n)}(\bm{\theta})}\\
        &+\cos({\theta_g})\left(\bra{\Psi^{(n)}(\bm{\theta})}\{\hat{H},\hat{\tau}_g^2\}\ket{\Psi^{(n)}(\bm{\theta})} - 2\bra{\Psi^{(n)}(\bm{\theta})}\hat{\tau}_g^2\hat{H}\hat{\tau}_g^2\ket{\Psi^{(n)}(\bm{\theta})}\right)\\
        &+\sin({\theta_g})\left(\bra{\Psi^{(n)}(\bm{\theta})}[\hat{H},\hat{\tau}_g]\ket{\Psi^{(n)}(\bm{\theta})} - \bra{\Psi^{(n)}(\bm{\theta})}\hat{\tau}_g[\hat{H},\hat{\tau}_g]\hat{\tau}_g\ket{\Psi^{(n)}(\bm{\theta})}\right)\\
        &+\cos({\theta_g})\sin({\theta_g})\bra{\Psi^{(n)}(\bm{\theta})}\hat{\tau}_g[\hat{H},\hat{\tau}_g]\hat{\tau}_g\ket{\Psi^{(n)}(\bm{\theta})} + \cos^2{\theta_g}\bra{\Psi^{(n)}(\bm{\theta})}\hat{\tau}_g^2\hat{H}\hat{\tau}_g^2\ket{\Psi^{(n)}(\bm{\theta})}\\
        &- \sin^2({\theta_g})\bra{\Psi^{(n)}(\bm{\theta})}\hat{\tau}_g\hat{H}\hat{\tau}_g\ket{\Psi^{(n)}(\bm{\theta})}.
    \end{split}
\eea
By apply the substitutions $\cos^2{\theta_g} = 1-\sin^2{\theta_g}$, $\sin^2{\theta_g} = \frac{1-\cos{2\theta_g}}{2}$ and $\cos{\theta_g}\sin{\theta_g} = \frac{\sin{2\theta_g}}{2}$, 
we obtain the expression
\bea
    \begin{split}
        E_g^{(n+1)}(\theta_g) &=\cos({\theta_g})\underbrace{\left(\bra{\Psi^{(n)}(\bm{\theta})}\{\hat{H},\hat{\tau}_g^2\}\ket{\Psi^{(n)}(\bm{\theta})} - 2\bra{\Psi^{(n)}(\bm{\theta})}\hat{\tau}_g^2\hat{H}\hat{\tau}_g^2\ket{\Psi^{(n)}(\bm{\theta})}\right)}_{a_1}\\
        &+\sin({\theta_g})\underbrace{\left(\bra{\Psi^{(n)}(\bm{\theta})}[\hat{H},\hat{\tau}_g]\ket{\Psi^{(n)}(\bm{\theta})} - \bra{\Psi^{(n)}(\bm{\theta})}\hat{\tau}_g[\hat{H},\hat{\tau}_g]\hat{\tau}_g\ket{\Psi^{(n)}(\bm{\theta})}\right)}_{b_1}\\
        &+\cos({2\theta_g})\underbrace{\left(\frac{1}{2}\bra{\Psi^{(n)}(\bm{\theta})}\hat{\tau}_g\hat{H}\hat{\tau}_g\ket{\Psi^{(n)}(\bm{\theta})}+\frac{1}{2}\bra{\Psi^{(n)}(\bm{\theta})}\hat{\tau}_g^2\hat{H}\hat{\tau}_g^2\ket{\Psi^{(n)}(\bm{\theta})}\right)}_{a_2}\\
        &+\sin({2\theta_g})\underbrace{\bra{\Psi^{(n)}(\bm{\theta})}\hat{\tau}_g[\hat{H},\hat{\tau}_g]\hat{\tau}_g\ket{\Psi^{(n)}(\bm{\theta})}}_{b_2}\\
        &+ \underbrace{\bra{\Psi^{(n)}(\bm{\theta})}\hat{H}\ket{\Psi^{(n)}(\bm{\theta})} - \bra{\Psi^{(n)}(\bm{\theta})}\{\hat{H},\hat{\tau}_g^2\}\ket{\Psi^{(n)}(\bm{\theta})}}_{c}\\
        &+\underbrace{\frac{3}{2}\bra{\Psi^{(n)}(\bm{\theta})}\hat{\tau}_g^2\hat{H}\hat{\tau}_g^2\ket{\Psi^{(n)}(\bm{\theta})} - \frac{1}{2}\bra{\Psi^{(n)}(\bm{\theta})}\hat{\tau}_g\hat{H}\hat{\tau}_g\ket{\Psi^{(n)}(\bm{\theta})}}_{c}.
    \end{split}
\eea

\section{Comparison between standard and efficient fermionic Rotoselect}
\begin{figure*}[t!]
    \centering
    \includegraphics[width=\textwidth]{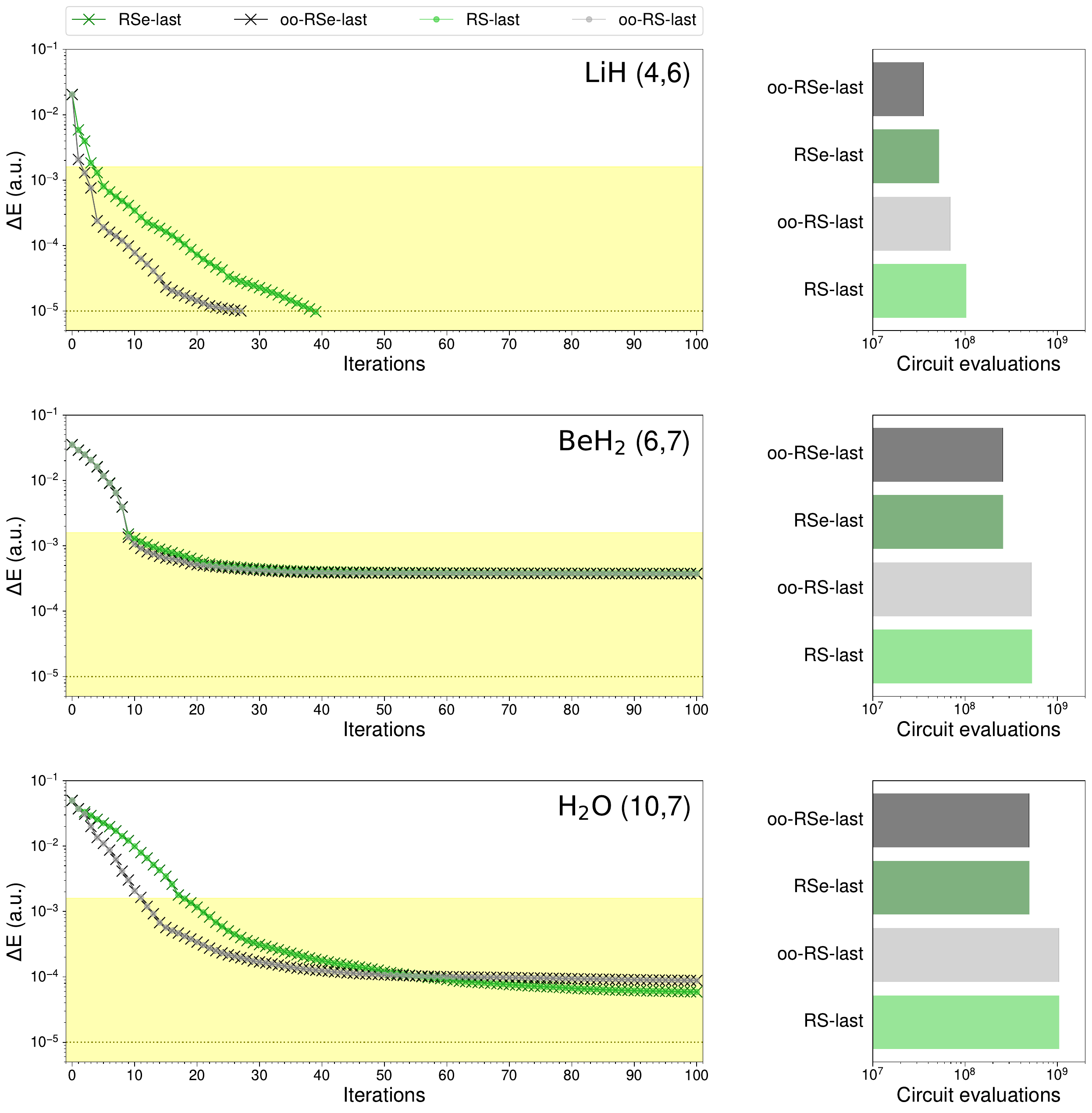}
    \caption{Optimization path and cost comparison of the RS and RSe algorithms in equilibrium LiH (4,6), \ce{BeH2} (6,7) and \ce{H2O} (10,7). Left column: energy convergence with the number of iterations of both the fixed orbitals (no prefix) and orbital-optimized ('oo' prefix) models. Right column: selection cost of each model in terms of number of noiseless circuit evaluations.}
    \label{fig:RSvsRSe-eq_geom}
\end{figure*}

\begin{figure*}[t!]
    \centering
    \includegraphics[width=\textwidth]{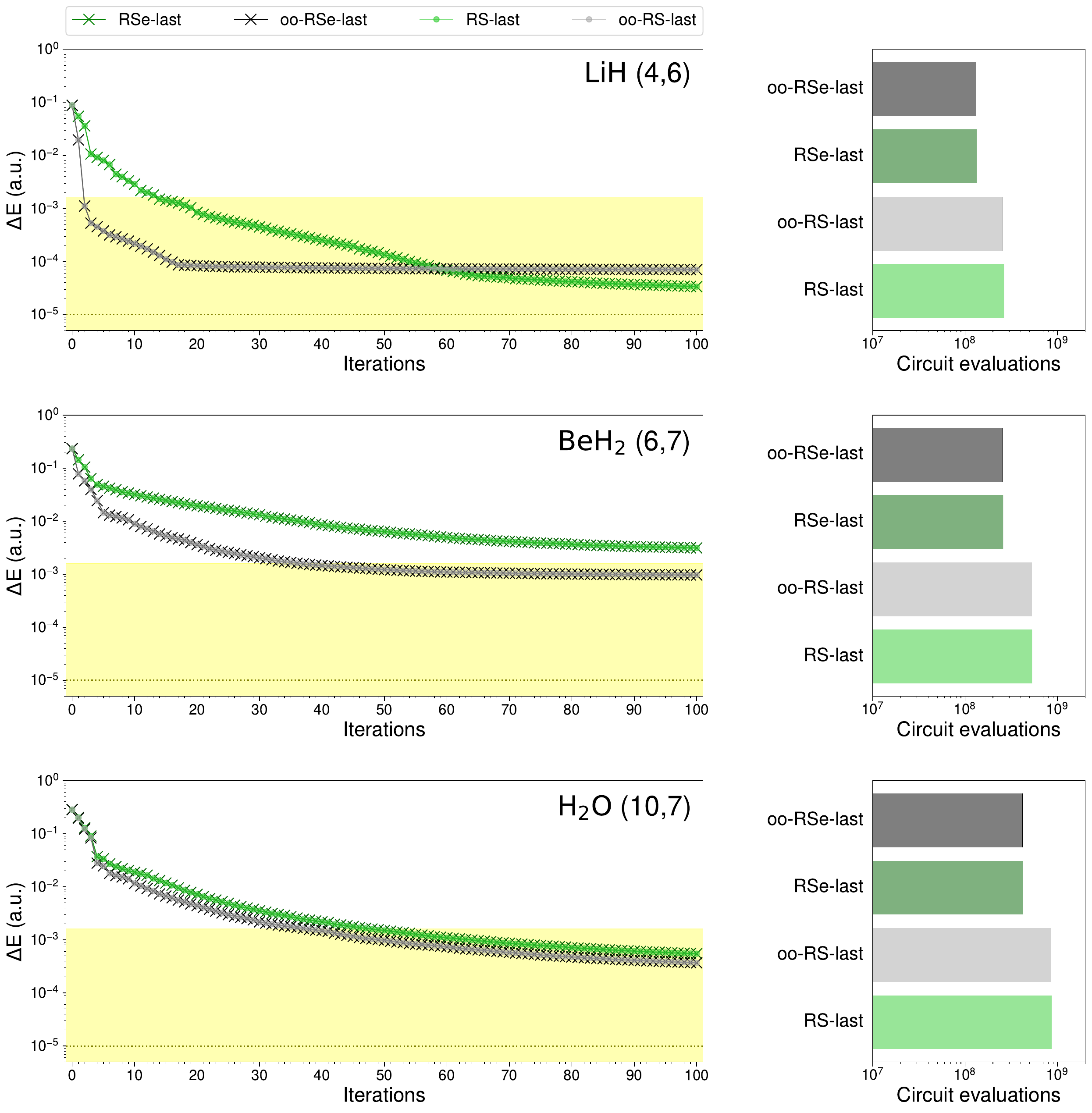}
    \caption{Optimization path and cost comparison of the RS and RSe algorithms in stretched LiH (4,6), \ce{BeH2} (6,7) and \ce{H2O} (10,7). Left column: energy convergence with the number of iterations of both the fixed orbitals and orbital-optimized models. Right column: selection cost of each model in terms of number of noiseless circuit evaluations.}
    \label{fig:RSvsRSe-str_geom}
\end{figure*}

In Fig. \ref{fig:RSvsRSe-eq_geom} and Fig. \ref{fig:RSvsRSe-str_geom} we compare the ansätze obtained using the efficient Rotoselect algorithm (RSe) and the current state-of-the-art Rotoselect algorithm (RS) as implemented in \cite{GGA_ADAPT_2025,ExcitationSolve_2025}. Here, we include only the models using the \textit{last} optimization strategy, in order to offer a comparison based solely on the selection step. The perfect equivalence between the RSe and RS models is evident from the comparison between the respective fixed-orbitals and orbital-optimized models. The comparison of the selection cost in the bar plots in the right columns of Fig. \ref{fig:RSvsRSe-eq_geom} and Fig. \ref{fig:RSvsRSe-str_geom} shows the (roughly) factor of two reduction in the number of circuit evaluations afforded by RSe compared to RS.

\section{Comparison between warm and zero start VQE optimization}
In this section, we provide a quantitative assessment of the effects of the \textit{warm start} and \textit{zero start} on the convergence and cost of the VQE optimization. In the zero-start models, the parameter of the appended operator is initialized to zero; in the warm-start models, such parameter is initialized to the value obtained through the RSe procedure. To ensure a fair comparison, we optimize the exact same ansatz, constructing it by sequentially appending operators from the sequence determined by the RSe-last model for the given system.

\begin{figure}[h]
    \centering
    \includegraphics[width=\textwidth]{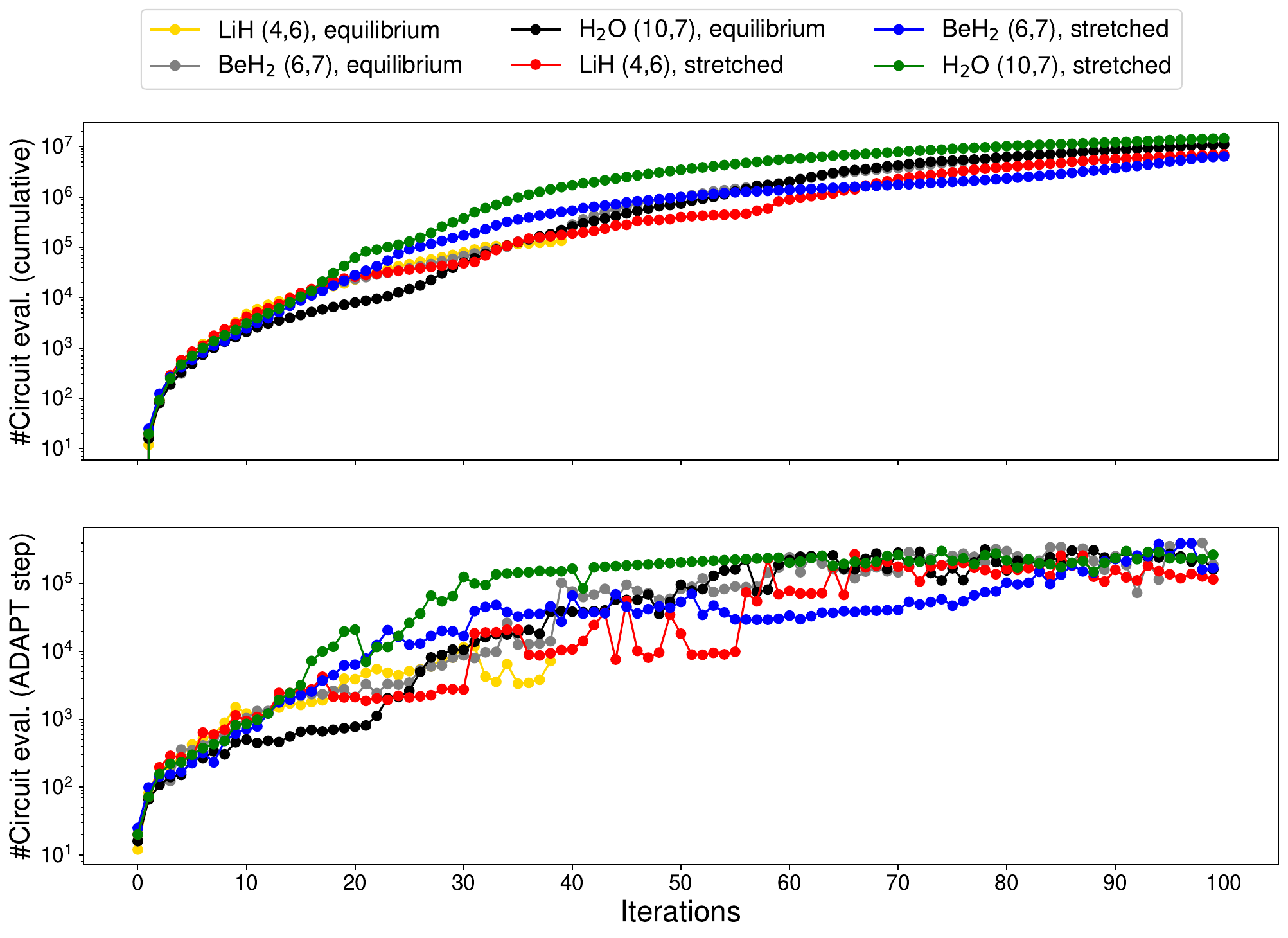}
    \caption{Cost (in terms of number of ungrouped circuit evaluation) of the VQE optimization of the last unitary starting from a zero parametrization for each ADAPT iteration. For each system, the ansatz is constructed using the same sequence of operators obtained by the RSe-last model. Upper panel: Cumulative cost as a function of the number of ADAPT iterations. Lower panel: Cost per ADAPT iteration.}
    \label{fig:zerovswarm_last}
\end{figure}

\subsection{Last optimization only}
First, we compare how the initialization approaches affect the \textit{last} optimization strategy. In this case, the warm-start strategy requires no additional VQE steps, as the Rotoselect procedure already delivers the parameter values corresponding to the minimum of the one-parameter cost function. This does not hold for the zero-start strategy, which requires a VQE optimization to obtain the optimal parameter. This extra cost is shown in \fig{fig:zerovswarm_last}. The upper panel displays the cumulative optimization cost over all ADAPT iterations, while the lower panel shows the cost for each ADAPT iteration step. It is evident that the zero start requires a substantially additional number of circuit evaluations (i.e., in the order of $10^5$ for LiH (4,6) and $10^6-10^7$ circuit evaluations for the other systems) compared to the warm-start strategy. Interestingly, the lower panel of \fig{fig:zerovswarm_last} shows that the cost per ADAPT iteration is not constant; instead, it grows steadily over the first thirty iterations before plateauing as more unitaries are appended to the ansatz. 
We note that the VQE optimisation and the optimal parameter found be RSe are the same.

\begin{figure}[h]
    \centering
    \includegraphics[width=\textwidth]{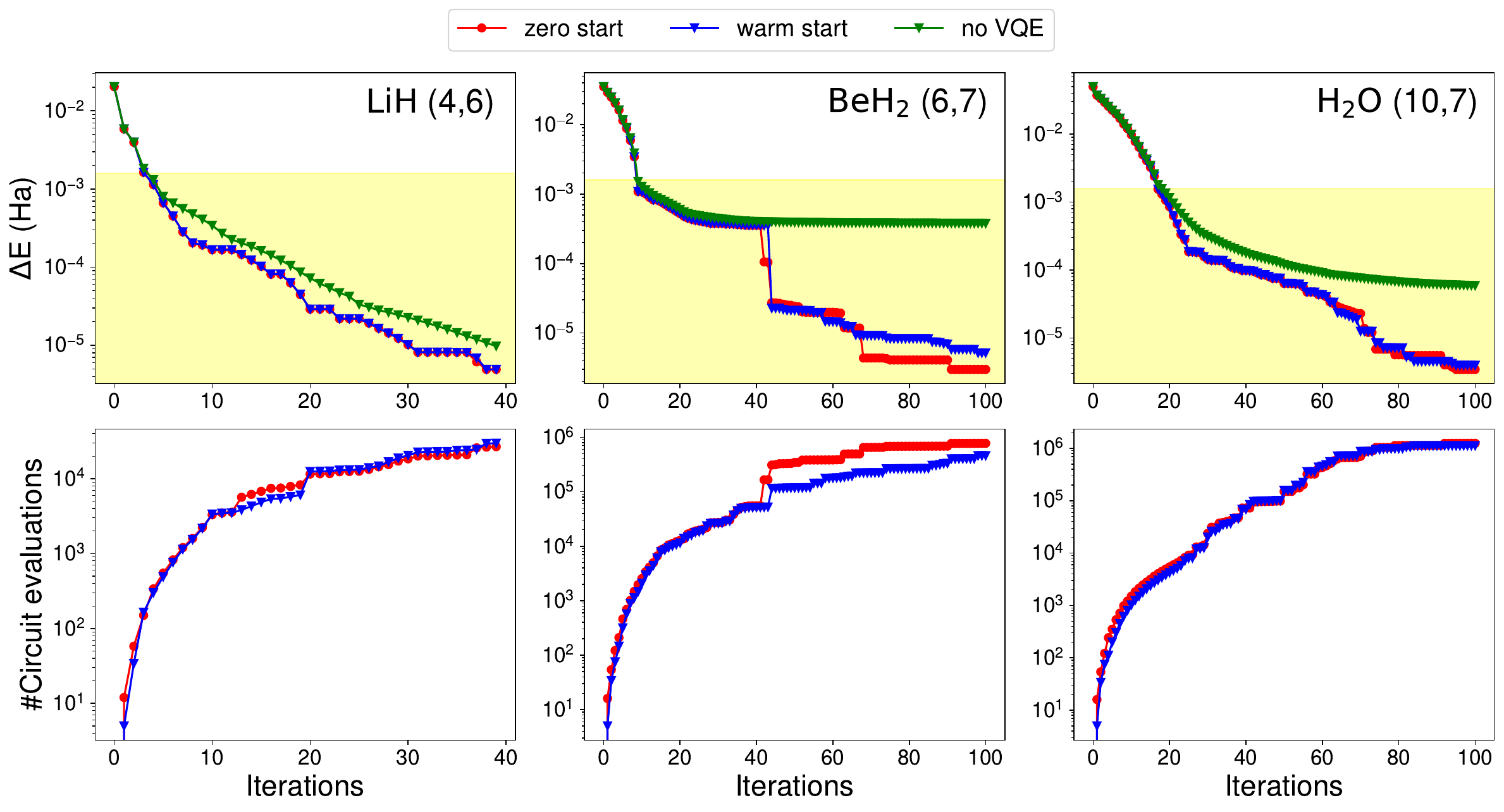}
    \caption{Comparison of zero-start and warm-start full VQE optimization of equilibrium LiH (4,6), \ce{BeH2} (6,7), and \ce{H2O} (10,7). Upper panels: Energy difference relative to the CASCI reference as a function of the number of ADAPT iterations. Lower panel: Cost (in terms of number of circuit evaluations) of warm start and zero start optimizations as a function of the number of ADAPT iterations.}
    \label{fig:zerovswarm_eq}
\end{figure}

\subsection{Full optimization}
\begin{figure}[h]
    \centering
    \includegraphics[width=\textwidth]{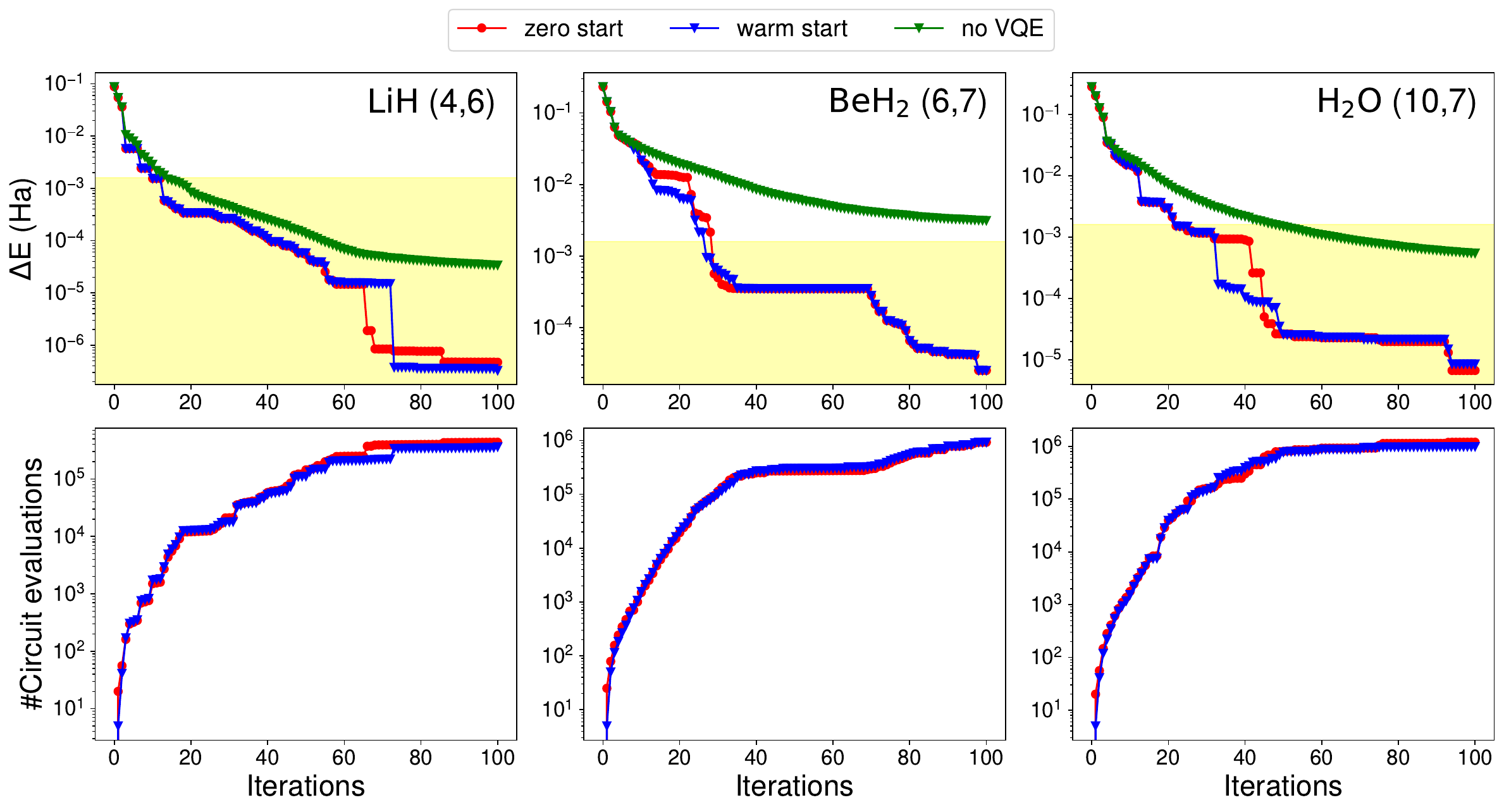}
    \caption{Comparison of zero-start and warm-start full VQE optimization of stretched LiH (4,6), \ce{BeH2} (6,7), and \ce{H2O} (10,7). Upper panels: Energy difference relative to the CASCI reference as a function of the number of ADAPT iterations. Lower panel: Cost (in terms of number of circuit evaluations) of warm start and zero start optimizations as a function of the number of ADAPT iterations.}
    \label{fig:zerovswarm_str}
\end{figure}

As a second step, we compare how the zero-start and warm-start affect the \textit{full} VQE optimization at each ADAPT iteration. The results of the analysis are shown in \fig{fig:zerovswarm_eq} and \ref{fig:zerovswarm_str} for the equilibrium and stretched systems, respectively. The upper panels show the energy convergence, while the lower panels show the circuit evaluation costs. Therein, the green line indicates the RSe-last reference without VQE optimization, while the red and blue lines show the full VQE optimization with a zero and warm start, respectively. 
In line with the conclusion of the main text, we see a clear impact of full VQE optimization (red and blue lines) on the energy convergence compared to no VQE optimization (green line). As expected, this effect is most pronounced for the highly correlated stretched test cases and has only a minor impact on equilibrium LiH.
Comparing the red (zero start) and blue (warm start) lines gives us a direct indication of the effect of warm start of the last parameter on the full VQE optimization. For all examples, the warm start is connected to a lower cost in the early iterations of the adaptive algorithm; however, as more layers are added, the early advantage of the warm start progressively fades as the optimization costs are dominated by re-optimizing previous layers. The final difference in energy and measurement costs is then dominated by the particular local minimum found rather than warm vs. zero start. Therefore, from the examples considered here, we conclude that the warm-start initialization of the newly added parameter does not lead to substantial advantage for full VQE optimization, neither in terms of the energy of the optimized ansatz, nor in terms of optimization cost.
\pagebreak
\bibliography{references}